\newcommand*\figpath{./}

\documentclass[fontsize=12pt]{article}

\usepackage{natbib}
\usepackage{graphicx}   
\usepackage[utf8]{inputenc} 
\usepackage[colorlinks=true, linkcolor=myred, citecolor=myred]{hyperref}         
\usepackage{listings}
\usepackage[outdir=./]{epstopdf}
\usepackage{amsmath,amsfonts,amssymb}
\usepackage{mathrsfs}
\usepackage{amsthm}
\usepackage{pdfpages}	
\usepackage{booktabs}	
\usepackage{times}            
\usepackage{colordvi}        
\usepackage{color}
\usepackage{xcolor}
\usepackage{multimedia}       
\usepackage{array,multirow,makecell}  
\usepackage{tabularx}
\usepackage{enumitem}
\usepackage{csvsimple}
\usepackage{comment}
\usepackage{float}
\usepackage{subcaption}
 \usepackage{lineno}

\usepackage{color}
\definecolor{myred}{rgb}{0.8,0.1,0.1}
\definecolor{mygreen}{rgb}{0.1,0.8,0.1}
\definecolor{darkgreen}{rgb}{0.1,0.4,0.1}
\definecolor{mygray}{gray}{0.93}

\usepackage{listingsutf8}

\lstdefinestyle{ffem}{
	framerule=0.6pt,
	texcl=true,
	basicstyle=\scriptsize\ttfamily\bfseries,
	keywordstyle=\color{blue}\bfseries,
	emphstyle=\color{blue}
	\bfseries,
	commentstyle=\color{darkgreen}\bfseries
	\itshape,
	stringstyle=\itshape,
	tabsize=4,
	frame=lines,
	numbers=none,
	numberstyle=\tiny,
	breaklines=true,
	showstringspaces=false,
	morekeywords={real,plot,border,mesh,label,buildmesh,adaptmesh,cmm,problem,int2d,int1d,fespace,func,string,on,dx,dy,cout,for,if,int,ofstream,ifstream},
	literate={?}{{\`a}}1 {�}{{\`e}}1 {�}{{\'e}}1,
	moredelim=[is][\color{blue}\footnotesize\bf]{|}{|},
	moredelim=[is][\color{darkgreen}\footnotesize\bf]{|+}{+|},
	morecomment=[l]{//},
	morestring=[s]{"}{"},
	morecomment=[s]{/*}{*/},
	escapeinside={@*}{*@},
	abovecaptionskip=0cm}

\usepackage[font=it,skip=\baselineskip]{caption}
\usepackage[normalem]{ulem}

\def\ds{\displaystyle}
\def\pl{\partial}

\renewcommand{\vec}[1]{\boldsymbol{#1}}

\newcommand {\R} {{\mathbb R}}

\newcommand{\ie}{{\em i.\thinspace{}e. }}

\newcommand{\eg}{{\em e.\thinspace{}g. }}

\newcommand{\ff}{FreeFem{\small +$\!$+}  }

\newcommand{\trap}{{\rm trap}}
\newcommand{\trunc}{{\rm trunc}}

\newcommand{\eff}{{\rm eff}}
\newcommand{\fit}{{\rm fit}}

\newcommand{\hb}{\hbar}
\newcommand{\agrad}{ A^t \nabla}
\newcommand{\TF} {{\hbox{\tiny TF}}}
\newcommand{\rtf}{\rho_{\hbox{\tiny TF}}}
\newcommand{\aho}{a_{\hbox{\footnotesize ho}}}
\newcommand{\omegap}{\omega_{\perp}}

\usepackage[hmargin={3cm,2.3cm},vmargin={3cm,3cm}]{geometry}

\def\LOGO{}
\font\fonteupmc=pagk at 10 true pt
\font\plutotgros=pagk at 12 true pt
\def\LAN{\plutotgros}
\def\upmc{\fonteupmc}

\def\entete{
	\vbox {
		\hbox to \hsize{%
			\hskip -1 cm\vbox{\LOGO}\hskip 0.45 true cm \vbox{\hbox{\LAN
					Laboratoire de math�matiques Raphael Salem} \vskip .2 true cm \hbox{\upmc
					Universit� de Rouen} \vskip .5 true cm
				\hbox{\upmc Avenue de l'Universit�, BP.12,
					76801 Saint-�tienne-du-Rouvray}}
			\hfill}
		{\vskip .9 true cm
			\hbox{\hskip -1 cm} \vskip .1 true cm
			\hbox{\hskip -1 cm}}}}

\title{\vspace{-1cm}Identification of vortices in quantum fluids: finite element algorithms and programs}

\author{Victor Kalt$^{1}$, Georges Sadaka$^{1}$, Ionut Danaila$^{1}$$^{*}$, Fr{\'e}d{\'e}ric Hecht$^{2}$}
\date{\small $^{1}$Universit{\'e} de Rouen Normandie, CNRS UMR 6085, Laboratoire de Math{\'e}matiques Rapha{\"e}l Salem,\\  \small Avenue de l'Universit{\'e}, BP 12, F-76801 Saint-{\'E}tienne-du-Rouvray, France \\
	\small $^{2}$Sorbonne Universit{\'e}, CNRS UMR 7598, Laboratoire Jacques-Louis Lions, F-75005, Paris, France. \\
	\small $^{*}$Corresponding author. Tel.: (+33) 2 32 95 52 50 \\
	}

\providecommand{\keywords}[1]
{
	\small	
	\noindent\textbf{\textit{Keywords---}} #1
}

\providecommand{\Email}[1]
{
	\small	
	\noindent\textbf{\textit{Email---}} #1
}

\begin{document}
\maketitle

\begin{abstract}
We present finite-element numerical algorithms for the identification of vortices in quantum fluids described by a macroscopic complex  wave function. 
Their implementation using the free software FreeFem++  is distributed with this paper as a post-processing toolbox that can be used to analyse numerical or experimental data. Applications for  Bose-Einstein condensates (BEC) and superfluid helium flows are presented.
Programs are tested and validated using either numerical data obtained by solving the Gross-Pitaevskii equation or experimental images of rotating BEC. Vortex positions are computed as topological defects (zeros) of the wave function when numerical data are used. For experimental images, we compute vortex positions as local minima of the  atomic density, extracted after a simple image processing. Once vortex centers are identified, we use a fit with a  Gaussian to precisely estimate vortex radius. For vortex lattices, the lattice parameter (inter-vortex distance) is also computed. The post-processing toolbox  offers a complete description of vortex configurations in superfluids. Tests for two-dimensional  (giant vortex in rotating BEC, Abrikosov vortex lattice in experimental BEC) and three-dimensional (vortex rings, Kelvin waves and quantum turbulence fields in superfluid helium) configurations show the robustness of the software. The communication with programs providing the numerical or experimental wave function field is simple and intuitive. The post-processing toolbox can be also applied for the identification of vortices in superconductors.
\end{abstract}

\\
\Email{victor.kalt@univ-rouen.fr, georges.sadaka@univ-rouen.fr, ionut.danaila@univ-rouen.fr,\\ frederic.hecht@sorbonne-universite.fr}

\noindent
{\bf Programm summary}\\
{\em Program Title:}  {FFEM\_postproc\_data\_2D, FFEM\_postproc\_data\_3D and FFEM\_postproc\_image}\\
{\em Catalogue identifier:}\\ 
{\em Program summary URL:}\\
{\em Program obtainable from:}\\
{\em Licensing provisions:} \\
{\em No. of lines in distributed program, including test data, etc.:} 81429 \\
{\em No. of bytes in distributed program, including test data, etc.:}  8 768 685\\
{\em Distribution format:} .zip\\
{\em Programming language:} \ff  (v 4.12) free software (www.freefem.org)\\
{\em Computer:} PC, Mac, Super-computer.\\
{\em Operating system:} Mac OS, Linux, Windows.\\
{\em Nature of problem:} The software is scoped to the identification of quantized vortices in 2D or 3D configurations described by a complex wave function.
Either simulation data obtained through numerical resolution of the Gross-Pitaevskii equation or 2D experimental images of Bose-Einstein condensates (BEC) can be used as input data. Examples of vortex identification in rotating BECs and superfluid helium are illustrated in the paper.\\
{\em Solution method:} We model the complex wave function using P1 (piece-wise linear) Galerkin triangular (in 2D) and tetrahedral (in 3D) finite elements. For simulation data, the zeros of the wave function are directly computed and the circulation of the velocity on  triangles sides is assessed. Vortices are obtained when a zero is found in a triangle with non-zero circulation. In 3D, the algorithm is applied on tetrahedron faces and the zero points are linked using mesh connectivity to form vortex lines. For experimental images, vortices are identified as local minima of the atomic density and extracted using the image contrast. \\
{\em Running time:} From seconds to minutes depending on the mesh resolution of the simulation/experimental data.\\


\section{Introduction}

The nucleation of quantized vortices, with fixed (quantized) circulation  and fixed core diameter is a striking feature of the two main known quantum fluids: superfluid helium and atomic  Bose-Einstein condensate (BEC). Liquid helium below the critical (lambda) temperature $T_\lambda =2.17K$ is  also called He II and is generally described as a  mixture of two fluids with independent velocity fields: a {\em normal} viscous fluid and an inviscid {\em superfluid}. In the superfluid component,  a large number of quantized vortices of the atomic size are nucleated and their complex interactions lead to a Quantum Turbulence (QT) state. The vortex line density plays an important role in the statistical description of QT and is usually  measured in experiments using second sound probes \citep{QT-experiment-2009-Donnelly}. Visualisations of vortices in superfluid helium are relatively recent and offer only a qualitative picture of vortex reconnections  
\citep{QT-experiment-2006-visu-Bewley}. Therefore, considerable theoretical and numerical efforts (see dedicated reviews or volumes \cite{QT-book-sreeni-2012,QT-review-2014-Barenghi-PNAS,QT-review-2017-tsubota-num}) were devoted to unravel properties of vortex interactions in He II. The Gross-Pitaevskii (GP) model has been extensively used  to explore properties of QT in an ideal setting containing only the superfluid \citep{Nore97a,Abid2003509,dan-2021-CPC-QUTE}. Even though the GP equation offers only a partial description of the complexity of superfluid helium, quantum vortex interactions are accurately described without any additional phenomenological models. 

The extraction of vortices from a GP numerical dataset providing values of the complex macroscopic wave function (order parameter) $\psi$ is thus an important tool to study properties of QT turbulence in superfluid helium. The main difficulty in this setting is the existence of a very large number of vortices with tangled shapes resulting from multiple reconnections. The identification of vortices is however simplified by the fact that vortices are evolving into a uniform background quantum flow of constant atomic density $n=|\psi|^2=n_0$. One the main applications of the algorithms and programs presented in this paper is the extraction of multiple vortex lines from a given complex wave function field $\psi$, resulting from numerical simulations of QT based on the GP model.

Quantized vortices have been extensively studied in BECs, since their existence provides an evidence of the superfluid nature of these systems. Different vortex generation methods were used in experimental BEC setups: based on the
drag on an object moving through the condensates \citep{BEC-phys-1999-jackson}, by rotating the trap confining the  atoms  \citep{BEC-physV-2000-Madison-b,BEC-physV-2001-haljan}, or by phase-engineering of topological defects 
\citep{BEC-physV-2002-imprint}.  Experimental observations were systematically supplemented with numerical simulations based on the GP equation, which is in this case (dilute BEC at zero temperature) the standard mathematical model.  Different types of vortices have been observed in experiments or numerical simulations of BEC: Abrikosov triangular vortex lattices \citep{BEC-physV-2003-engels}, bent vortex lines of "U" or "S" shapes \citep{BEC-physV-2002-Rosenbusch-a,dan-2003-aft}, giant vortices \citep{BEC-physV-2004-bretin,dan-2004-aft,dan-2005} or more exotic shapes of vortices (double rings, hopfions, stars, etc.) \citep{BEC-BdG-2015-Panos,dan-2004-cras}. Experimental images of quantized vortices in BEC are typically obtained by switching-off the magnetic trap and imaging (after a time of flight) the absorption of a resonant laser beam propagating along the $z$-axis. The obtained pictures, showing atomic density integrated along the $z$-axis, are then used to extract the size and the average position of the condensate \citep{BEC-physV-2002-Rosenbusch-b}. Note that vortex cores are larger in BEC ($5\cdot 10^{-7} m$) than in superfluid helium ($10^{-10} m$) and thus more easily observable. Note also that recent experimental and theoretical studies \citep{QT-experiment-2014-Kwon,QT-experiment-2016-Navon} reported different possible routes to QT in BEC.

Another application of the algorithms and programs presented in this contribution is the possibility to accurately post-process BEC experimental images, as well as wave function fields provided by GP numerical simulations of BEC. The difficulty in the BEC case is that vortices are topological defects (zeros of the wave function) on the top of a non-constant atomic density that also vanishes outside the condensate. This background density depends on the BEC setting (trapping potential, rotation) and is taken into account in the algorithm extracting vortices.  The post-processing thus separates the non-condensed cloud from the condensate, extracts the background atomic density to isolate vortices and finally fits each individual vortex with an anisotropic Gaussian to compute vortex radius. 

Even though the present toolbox can be easily used with experimental images of quantum systems with vortices (superfluids, superconductors), the main intended application is the link with numerical systems providing a 2D or 3D field of the complex wave function $\psi$.  Numerical results of vortices in superfluids are generally presented using contours of the atomic density  in 2D simulations or iso-surfaces of low atomic density in 3D simulations. Precise methods for the tracking of vortices in complex fields were developed by \cite{villois2016vortex} using a Newton-Raphson method to determine the zeros of the wave function combined with Fourier interpolation for the evaluation of the wave function outside grid points. \cite{caliari2017postproc} suggested  another method adapted to spectral Fourier computations, based on non-equispaced Fast Fourier Transform to truncate Fourier series at arbitrary points and finally identify vortex line positions during vortex reconnection. 
A fast real-time visualisation technique of vortices in QT simulations of high resolution simulations was developed by \cite{liu2020vectorizing} using a graph-based method.

\pagebreak
The advantage of using the present toolbox is to use generic unstructured grids with finite elements in 2D and 3D that can be subsequently refined around vortices. Adapted mesh provides accurate identification of vortex centers, vortex lines description and vortex radius computation. These quantities are important in assessing physical and mathematical theories on vortex dynamics or QT in BEC or superfluid helium.  A great number of  software packages for solving different forms (stationary, time-dependent, with non-local or high-order interactions)  of the GP equation exist in the CPC Program Library. Spatial discretizations are generally based on spectral \citep{BEC-CPC-2007-dion-cances,BEC-CPC-2013-Caliari,BEC-CPC-2014-antoine-duboscq,BEC-CPC-2015-antoine-duboscq}, finite-elements \citep{BEC-CPC-2016-FEM,dan-2016-CPC} or finite-difference \citep{BEC-CPC-2009-Muruganandam,BEC-CPC-2012-Vudragovic,BEC-CPC-2013-Caplan,BEC-CPC-2014-simplectic,BEC-CPC-2014-Hohenester,BEC-CPC-2019-rotating} methods. Provided programs are written in Fortran  \citep{BEC-CPC-2007-dion-cances,BEC-CPC-2009-Muruganandam}, C \citep{BEC-CPC-2012-Vudragovic,BEC-CPC-2013-Caplan}, Matlab \citep{BEC-CPC-2013-Caliari,BEC-CPC-2013-Caplan,BEC-CPC-2014-antoine-duboscq,BEC-CPC-2014-Hohenester,BEC-CPC-2015-antoine-duboscq}, FreeFem++ \citep{dan-2016-CPC} or C and Fortran with OpenMP \citep{BEC-CPC-2019-rotating}.
All these programs could benefit from the use of the present post-processing toolbox that can be easily interfaced with any type of space discretization by writing the output complex wave function fields in a standard finite-element format (examples of how to write data are provided in Sect. \ref{sec-desc-prog}).

The organization of the paper is as follows. In Sect. \ref{sec-gpe}  we introduce the GPE model and the characterisation of quantized vortices. Sections \ref{sec-2D-data} and \ref{sec-3D} present the main algorithms used for the identification of vortices in GP simulations of BECs for 2D and 3D configurations, respectively. Section \ref{sec-2D-images} is dedicated to the identification of vortices in experimental images. 
Programs architecture and main parameters of the toolbox are described in Sect. \ref{sec-desc-prog}. We present various 2D (Abrikosov vortex lattice, giant vortex) and 3D (bent vortex, vortex knot, Kelvin waves and vortices in QT) benchmarks used for the validation of our codes in Sect. \ref{sec-example}. Finally, the main features of the toolbox and its possible extensions are summarized in Sect. \ref{sec-conclusions}.


\section{The Gross-Pitaevskii model and quantized vortices}\label{sec-gpe}

\subsection{The Gross-Pitaevskii equation} \label{sec-GP}

The Gross-Pitaevskii equation describes the time-space evolution of the macroscopic wave function of a dilute gaseous BEC in the limit of zero temperature. 
For a  BEC of $N$ atoms confined in a trapping potential $V_\text{trap}({\vec x})$ and rotating with angular velocity ${\vec \Omega}$, the GP equation is:
\begin{equation}
\label{eq-GP}
i\hb\frac{\partial \psi}{\partial t} = -\frac{\hb^2}{2m} \nabla^2 \psi + V_\trap \psi + g |\psi|^2 \psi - {\vec \Omega} \cdot {\vec {\mathcal L}}(\psi),
\end{equation}
where ${\vec {\mathcal L}}$ is the angular momentum, $\hb$ the reduced Planck constant and $m$ the atomic mass. The constant in front of the non-linear term depends on the scattering length $a_s$,  as $g = \frac{4\pi\hbar^2 a_s}{m}$. The atomic density $n({\vec x})=|\psi({\vec x})|^2$ vanishes outside the condensate  and thus natural boundary conditions for \eqref{eq-GP} are homogeneous Dirichlet conditions ($\psi(\vec{x}) \rightarrow 0$ as $\vec{x} \rightarrow \infty$). When Eq. \eqref{eq-GP} is used to describe superfluid helium, the trapping potential and the angular velocity are set to zero and periodic boundary conditions are assumed. 
Considering rotations along the $z$-axis (\ie ${\vec \Omega}= \Omega\, {\vec k} )$ implies that only the $z$-component of the angular momentum appears in Eq. \eqref{eq-GP} and the rotation term can be presented as \citep{dan-2010-SISC,dan-2010-JCP,dan-2017-SISC}:
\begin{equation}
\label{eq-GP-Lz}
{\vec \Omega} \cdot {\vec {\mathcal L}}(\psi) = \Omega {\mathcal L}_z \psi =  i \hb \Omega \,\left(y\frac{\pl \psi}{\pl x} - x\frac{\pl \psi}{\pl y} \right) = i \hb \Omega\, \agrad \psi, \quad \mbox{with}\quad
A^t = (y , -x , 0).
\end{equation}
The following general {\em quadratic $\pm$ quartic} form of the trapping potential describes most of the existing theoretical and experimental studies of rotating BECs \citep{dan-2016-CPC}:
\begin{equation}
\label{eq-GP-trap-V}
V_\trap(x,y,z)=\frac{m}{2}\left(
\omega_{x}^2 {x}^2 + \omega_y^2 {y}^2 + \omega_z^2 {z}^2 \right)
+ U_2 \left(\frac{r}{w_2}\right)^2  
+ U_4 \left(\frac{r}{w_4}\right)^4, r^2 = x^2 + y^2,
\end{equation}
where $\omega_{x}, \omega_{x}, \omega_{x}$ are trapping frequencies of the classical harmonic ($\sim r^2$)  trapping potential and $U_2, U_4, w_2, w_4$ are characteristics of the detuned laser beam adding the quartic part ($\sim r^4$) of the potential and a supplementary (positive or negative) quadratic term \citep{BEC-physV-2004-bretin}. 
The conservation of the number of atoms $N$ inside the condensate is expressed as:
\begin{equation}
\label{eq-GP-cons}
\int_{\R^3} \left|\psi({\vec x})\right|^2\, d{\vec x} = N.
\end{equation}

\pagebreak
\subsection{The Thomas-Fermi approximation} \label{sec-TF}

The Thomas-Fermi regime is characterized by strong interactions (the kinetic energy is negligible compared to the interaction energy). In absence of rotation, the Thomas-Fermi density is classically \citep{BEC-book-2003-pita} obtained from Eq. \eqref{eq-GP} by considering stationary solutions $\psi (\vec{x},t) = \phi(\vec{x})\exp(-i \mu t / \hbar)$ and neglecting the Laplacian term (corresponding to kinetic energy):
\begin{equation}
V_\trap(\vec{x}) + g |\phi|^2 = \mu.
\label{eq-TF}
\end{equation}
The corresponding Thomas-Fermi atomic density then takes a simple form:
\begin{equation}
n_\TF (\vec{x}) = |\phi_\TF|^2 = \frac{\mu-V_\trap(\vec{x})}{g}.
\label{eq-TF-n}
\end{equation}
The constant $\mu$ is the chemical potential and its value can be calculated by combining Eqs. \eqref{eq-TF-n} and \eqref{eq-GP-cons}. If $\Omega \neq 0$, an equivalent Thomas-Fermi approximation can be used to approximate the density of the rotating condensate by considering \eqref{eq-TF-n}, but with an effective trapping potential:
\begin{equation}
\label{eq-GP-Veff}
V^\eff_\trap = V_\trap - \frac{1}{2} m \Omega^2 r^2, \quad r^2 =x^2 + y^2.
\end{equation}
The effective potential is the original potential diminished by the centrifugal term \citep{BEC-physV-1999-Stringari-TF,dan-2016-CPC}. In the case of superfluid helium ($V_\trap=\Omega=0$)  the Thomas-Fermi density is constant $n_\TF=n_0$.

\subsection{Quantized vortices} \label{sec-vortex}

Using the Madelung transformation, the wave function can be also presented as:
\begin{equation}
\psi(\vec{x},t) = \sqrt{n(\vec{x},t)} e^{iS(\vec{x},t)},
\end{equation} 
with $n$ the atomic density and $S$ the phase. The mass density $\rho$ and the velocity $\vec{v}$ of the superfluid are defined as:
\begin{align}
\rho(\vec{x},t) &= m\, n(\vec{x},t) = m\, |\psi(\vec{x},t)|^2,\\
\vec{v}(\vec{x},t) &= \frac{\hbar}{m}\nabla S(\vec{x},t) = \frac{\hbar}{\rho} \frac{\psi^*\nabla\psi - \psi \nabla\psi^*}{2i}.
\label{eq-vel}
\end{align}
For a flow of non-vanishing density ($\rho \neq 0$), we infer from \eqref{eq-vel} that the superfluid is irrotational:
\begin{equation}
\nabla \times \vec{v} = \frac{\hbar}{m} \nabla \times \nabla S =0.
\label{eq-vel-rot}
\end{equation}
The lines along which $\rho=0$ define topological defects, known as {\em quantized vortices}. It results from \eqref{eq-vel} that  the  velocity  is singular at the vortex center. The Madelung transformation becomes singular when vortices are present in the flow. For a  review of mathematical problems related to the Madelung transformation in presence of quantum vortices, see \cite{BEC-math-2013-carles-NL}. 

The circulation $\Gamma$  along a closed regular path $C$ surrounding a simply connected domain $S$ is:
\begin{equation}\label{eq-circ}
\Gamma = \oint_C \vec{v} \cdot d\vec{l} = \frac{\hbar}{m}\oint_C \nabla S \cdot d\vec{l}.
\end{equation}
If $S$ is a simply connected domain without vortices, using property \eqref{eq-vel-rot} and Stokes' theorem results in $\Gamma=0$. If a vortex is present in the domain $S$, the curl of the velocity expressed by  Eq. \eqref{eq-vel-rot}  becomes a Dirac function and a phase-defect line starting from the vortex center exists. Since $\psi$ must be a single-valued function, the integrated change in the phase must be a multiple of $2\pi$, which implies that:
\begin{equation}
\Gamma = \frac{\hbar}{m} 2\pi \kappa  =  \frac{h}{m}  \kappa ,
\end{equation}
where $\kappa$ is an integer representing the winding number of the vortex. It is important to note that vortex solutions are not singular solutions to the GP equation \eqref{eq-GP}. Indeed, a straight-line vortex solution $\Psi_{v}$ can be obtained  \citep{BEC-book-2003-pita} using cylindrical coordinates $(z, r, \varphi)$ as:
\begin{equation}
\label{eq-vortex-psi}
\Psi_{v} = \sqrt{n_0} \, f(r)\, e^{i \kappa \varphi}.
\end{equation}
The asymptotic behaviour of this solution near the origin ($r=0$) is known \citep{BEC-math-1990-neu}:
\begin{equation}
\label{eq-vortex-as}
f(r) \sim  r^{|\kappa|} + {\cal O} (r^{|\kappa|+2}), \quad r \rightarrow 0.
\end{equation}
The region near the vortex center where the density is varying in a significant way defines the vortex core. The radius of the vortex is of the order of the healing length $\xi$  \citep{BEC-book-2003-pita}:
\begin{equation}
\xi = \frac{\hbar}{\sqrt{2 m g |\psi|^2}} = \frac{1}{\sqrt{8 \pi a_s |\psi|^2}}.
\label{eq-rho0-xi-2}
\end{equation}

This description of quantized vortices offers a very simple practical way to extract vortices from a GP wave function field. Vortices are thus identified in the present post-processing as points in 2D or lines in 3D characterized by the following two properties:
\begin{itemize}
	\item The density at the vortex center is 0.
	\item The circulation of the velocity on a closed path around the center is non zero.
\end{itemize}

\subsection{Scaling} \label{sec-scaling}

Numerical simulations are usually based on the non-dimensional form of the GP equation \eqref{eq-GP}. By setting
\begin{equation}
\label{eq-scal-u}
{\vec{x}} \rightarrow \frac{\vec x}{\aho}, \quad u = \frac{\psi}{\sqrt{N} \, \aho^{-d/2}},
\end{equation}
where $d=2$ or $3$ is the dimension of the space and $\aho = \sqrt{\frac{\hb}{m\omegap}}$  the harmonic oscillator length defined with respect to a reference trapping frequency $\omegap=\min(\omega_x,\omega_y)$, the dimensionless form of the GP equation becomes:
\begin{equation}
\label{eq-scal-GP}
i \frac{\partial u}{\partial t} = -\frac{1}{2} \nabla^2 u + C_\text{trap} u + C_g |u|^2 u - i C_\Omega \agrad u.
\end{equation}
Coefficients in \eqref{eq-scal-GP} are defined as:
\begin{equation}
\label{eq-scal-Cg}
C_\Omega = \displaystyle \left(\frac{\Omega}{\omegap}\right), \quad C_g =  \frac{4\pi Na_s}{\aho} \quad \text{(in 3D)}, \quad C_g=\beta \quad \text{(given in 2D)}.
\end{equation}
The dimensionless effective potential corresponding to \eqref{eq-GP-Veff} and \eqref{eq-GP-trap-V} is : 
\begin{equation}
\label{eq-scal-trap-V}
V^\eff_\trap  = \frac{1}{2} \left(a_x x^2 + a_y y^2+ a_z z^2 + a_4 (x^2 + y^2)^2\right),
\end{equation}
with dimensionless coefficients \citep{dan-2016-CPC}:
\begin{equation}
\label{eq-scal-trap-ax}
\left\{
\begin{array}{lll}\vspace{0.2cm}
\displaystyle a_x= \left(\frac{\omega_x}{\omegap}\right)^2- 
\left(\frac{\Omega}{\omegap}\right)^2 +
2\left(\frac{U_2}{m \omegap^2 w_2^2}\right), & &\\ \vspace{0.2cm}
\displaystyle a_y= \left(\frac{\omega_y}{\omegap}\right)^2- 
\left(\frac{\Omega}{\omegap}\right)^2+ 
2\left(\frac{U_2}{m \omegap^2 w_2^2}\right),&\\ \vspace{0.2cm}
\displaystyle a_z= \left(\frac{\omega_z}{\omegap}\right)^2,\quad
\displaystyle a_4={2}\left(\frac{U_4\, \aho^2}{m \omegap^2 w_4^4}\right).&
\end{array}
\right.
\end{equation}
Non-dimensional forms \eqref{eq-scal-trap-V} and \eqref{eq-scal-trap-ax} are used to obtain analytical forms for the Thomas-Fermi approximation \eqref{eq-TF-n}:
\begin{equation}
\label{eq-app-TF}
\rtf = \mod{u}^2= \left(\frac{\rho_0 - 2{V}^\eff_\trap}{2 C_g}\right)_+.
\end{equation} 
Analytical expressions for $\rho_0$ are presented in detail by \cite{dan-2016-CPC} for trapping potentials of the form $quartic$ $\pm$ $quadratic$. Note that the non-dimensional healing length \eqref{eq-rho0-xi-2} becomes in the Thomas-Fermi approximation:
\begin{equation}
\frac{\xi}{\aho} = \frac{1}{\sqrt{2 C_g |u|^2}} \Longrightarrow   \frac{\xi_{\min}}{\aho}=\frac{1}{\sqrt{\rho_0}}.
\label{eq-rho0-xi-2-adim}
\end{equation}

\section{Vortex identification in 2D BECs}\label{sec-2D-data}

\begin{figure}
	\centering
	\begin{subfigure}[b]{0.48\textwidth}
		\centering
		\includegraphics[width=\textwidth]{\figpath/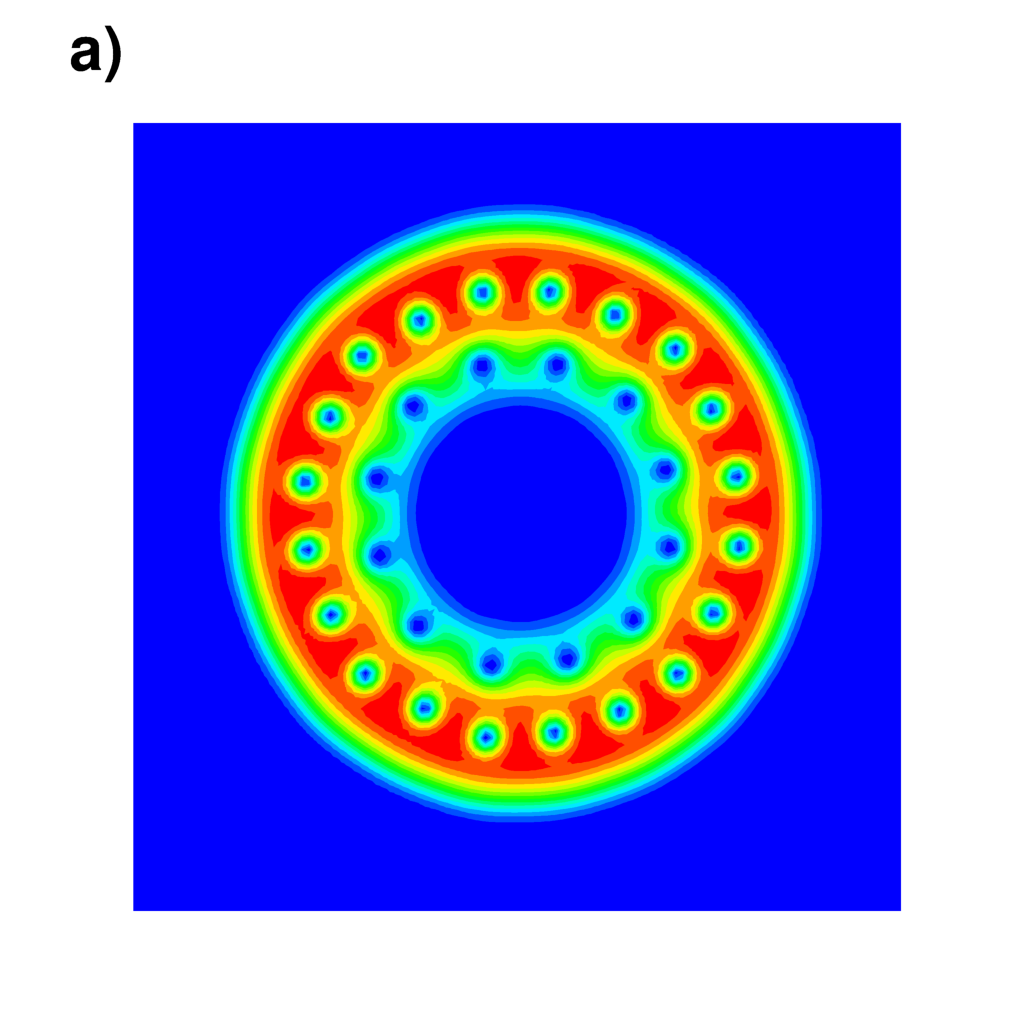}
		\label{fig:quartic_a}
	\end{subfigure}
	\hfill
	\begin{subfigure}[b]{0.48\textwidth}
		\centering
		\includegraphics[width=\textwidth]{\figpath/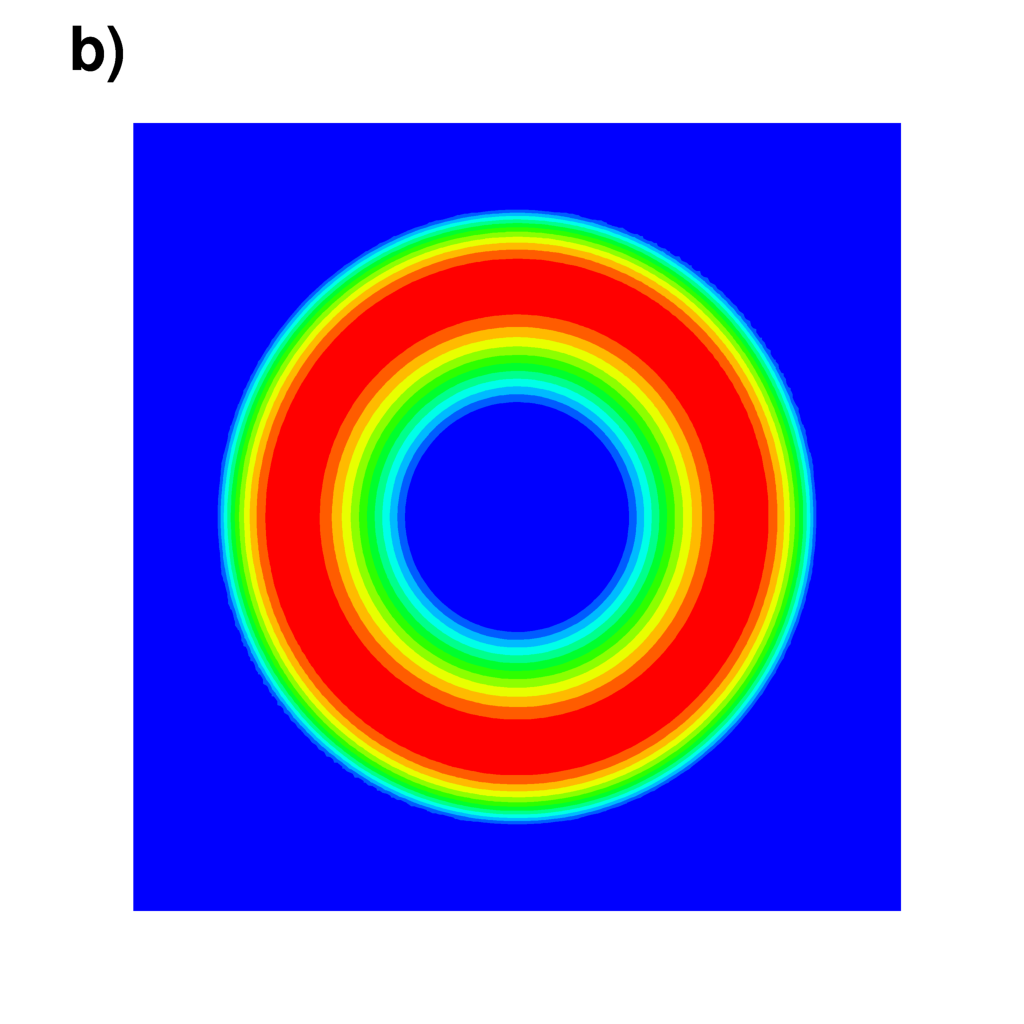}
		\label{fig:quartic_b}
	\end{subfigure}
	\begin{subfigure}[b]{0.48\textwidth}
		\centering
		\includegraphics[width=\textwidth]{\figpath/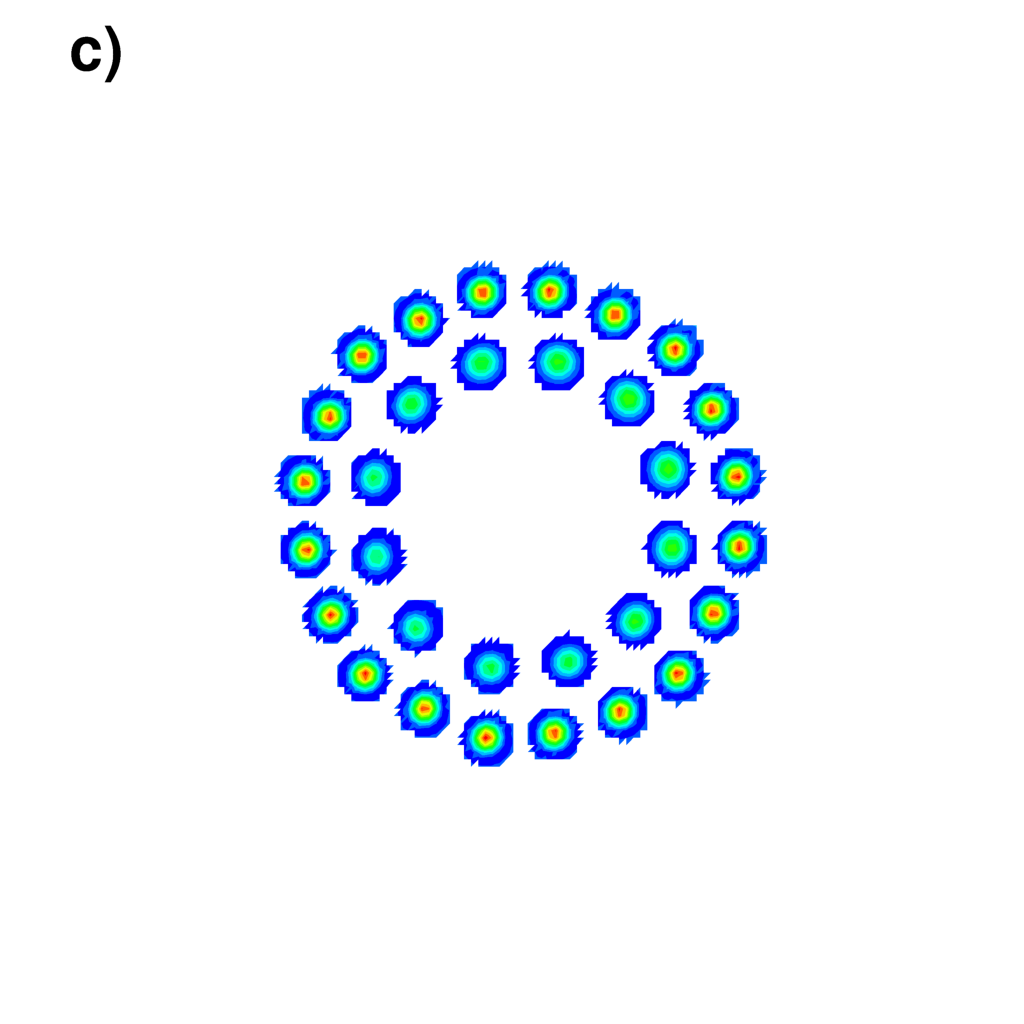}
		\label{fig:quartic_c}
	\end{subfigure}
	\hfill
	\begin{subfigure}[b]{0.48\textwidth}
		\centering
		\includegraphics[width=\textwidth]{\figpath/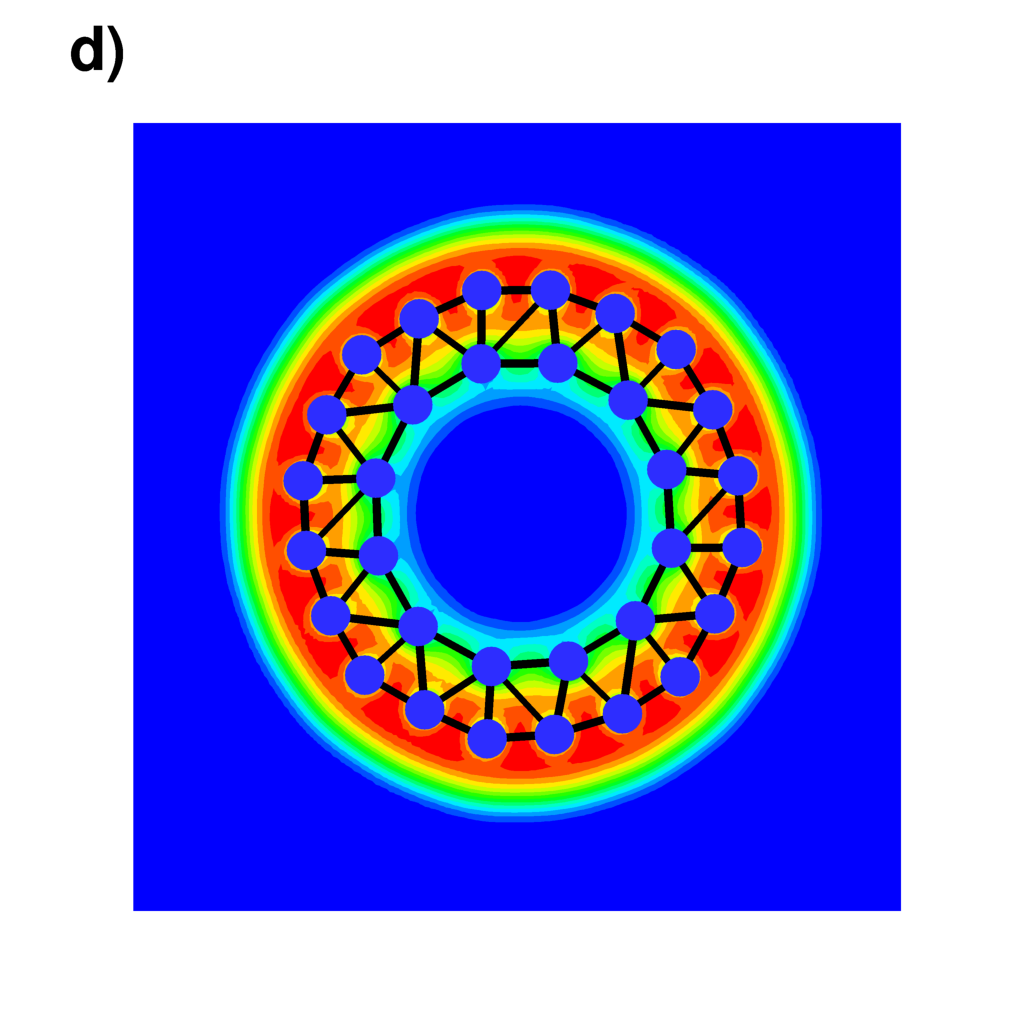}
		\label{fig:quartic_d}
	\end{subfigure}
	\caption{Illustration of vortex identification steps in 2D. Data from numerical simulation of a fast rotating BEC with a {\em quartic-minus-quadratic} trapping potential \citep{dan-2016-CPC}. a) Initial density $\rho$, b) Thomas-Fermi density $\rho_\TF$, c) identified vortex zones used to localize vortex centers and for the fitting with a Gaussian, d) mesh of the identified vortex lattice (blue points) on top of the initial data.}
	\label{fig-2D-quartic}
\end{figure}

The vortex identification process in 2D is decomposed into two parts: the localization of the vortex center and the computation of vortex characteristics. For the latter step, the vortex core radius (obtained through a fitting procedure) and the parameter of the Abrikosov vortex lattice (when identified) are computed. The initial data is the complex wave function $u$ obtained from numerical simulations and represented as a P1 (piece-wise linear) finite-element function on a triangular  mesh $Th$. 
We illustrate the algorithms using numerical data reported by \cite{dan-2016-CPC} for a fast rotating BEC with a {\em quartic-minus-quadratic} trapping potential. The initial data is shown in Fig. \ref{fig-2D-quartic}(a). A giant vortex is formed in the middle of the condensate, surrounded by a lattice of individual vortices. The purpose of the post-processing is to identify individual vortices and compute their characteristics. 

\subsection{Vortex localization}\label{sec-vortex-localization}

The main difficulty introduced by this case is that the wave function is vanishing inside the giant vortex and also outside the condensate. These zones have to be removed before searching the zeros of the wave function corresponding to vortices.  There are two possibilities depending on the availability of numerical parameters used in simulations.
\begin{itemize}
	\item If  parameters \eqref{eq-scal-trap-ax} of the trapping potential are known, we use formulae presented by \cite{dan-2016-CPC} to compute $\rho_{0}$ necessary to estimate the background density following the Thomas-Fermi approximation \eqref{eq-app-TF}.  For the case presented in this section
	the trapping potential 	\eqref{eq-scal-trap-V} is of the form ${V}^\text{eff}_\trap= \frac{1}{2}\left(a_2 r^2 + a_4 r^4\right)$ and we use the following algorithm:
	\begin{equation}
	\begin{array}{l}
	--> \mbox{compute $\rho_{0}$}\\ \vspace{0.2cm}
	\left[
	\begin{array}{l}\vspace{0.2cm}
	\text{Compute} \quad C_\text{S} = 2 C_g,\\
	\mbox{$\bullet$ if $a_2 <0$ and $\displaystyle a_4 < \sqrt{\frac{\pi a_2^3}{6 C_\text{S}}}$} \\ \vspace{0.2cm}
	\displaystyle \rho_{0} = \frac{1}{4 a_4}\left[\left(\frac{6 a_4^2}{\pi} C_\text{S}\right)^{2/3} - a_2^2\right] \\ \vspace{0.2cm}
	\mbox{$\bullet$ else}\\  \vspace{0.2cm}
	\mbox{calculate the root $\eta>0$ of:}\\ \vspace{0.2cm}
	\displaystyle f(\eta)=4 a_4 \eta^3 + 3 a_2 \eta^2 - \frac{6}{\pi} \left(C_\text{S} \right) = 0,\\ \vspace{0.2cm}
	\mbox{and then calculate}\\
	\displaystyle \rho_{0} = a_2 R^2 + a_4 R^4
	\end{array}
	\right.\\
	--> \mbox{compute the maximum radius of the condensate}\\ \vspace{0.2cm}
	\displaystyle R_+ = \left(\frac{-a_2 + \sqrt{a_2^2+4\rho_{0} a_4}}{2 a_4}\right)^{1/2}.
	\end{array}
	\label{eq-TF-detail}
	\end{equation}
	
	At the end of this step, mesh triangles where $\rho_\TF < 0$ are removed.
	
	\item If the analytical expression of the Thomas-Fermi approximation is not known, we use to identify the border of the condensate an iso-value line of very low density $\rho_{\min}$. A typical value is   $\rho_{\min}=5\cdot 10^{-5}$, but this threshold can be adjusted depending on the type of the scaling used in the numerical simulation. Triangles  where  $\ds \rho < \rho_{\min}$ are removed at the end of this step.
\end{itemize}

We can proceed now with vortex center localisation.
The location of vortices can  be obtained through a direct application of their defining properties mentioned at the end of Sect. \ref{sec-vortex}. For each triangle $T$ of the mesh, we search whether it  contains a zero of the wave function or not. We denote by $u_r=\Re(u)$ and $u_i=\Im(u)$ the real and imaginary part of the wave function, respectively. Let $P_0$, $P_1$, $P_2$ be  the vertices of the triangle. A necessary condition for the presence of a zero of a P1 complex function inside a triangle is that both real and imaginary part must change sign inside the triangle:
\begin{align}\label{eq-zero-tri}
&\min(u_r(P_0),u_r(P_1),u_r(P_2)) < 0,\\
&\min(u_i(P_0),u_i(P_1),u_i(P_2)) \, \,< 0,\\
&\max(u_r(P_0),u_r(P_1),u_r(P_2)) > 0,\\
&\max(u_i(P_0),u_i(P_1),u_i(P_2))  \, \,> 0.
\end{align}
The coordinates of the zero in barycentric coordinates are then computed as:
\begin{align}\label{eq-zero-point}
x_v &= \frac{\det(u(P_0),0,u(P_2))}{\det(u(P_0),u(P_1),u(P_2))},\\
y_v &= \frac{\det(u(P_0),u(P_1),0)}{\det(u(P_0),u(P_1),u(P_2))}.
\end{align}
If a zero is found inside the triangle, we compute the circulation of the velocity on triangle sides to assess whether a vortex is present. After discretizing Eq.  \eqref{eq-circ}, we use the following formula to compute circulation and corresponding winding number:
\begin{equation}\label{eq-circ-tri}
\kappa = \frac{1}{2\pi} \Im\left(\log\left(\frac{u(P_1)}{u(P_0)}\right) + \log\left(\frac{u(P_2)}{u(P_1)}\right) + \log\left(\frac{u(P_0)}{u(P_2)}\right)\right),
\end{equation}
where $\Im$ denotes the imaginary part.
A winding number $\kappa = \pm 1$ indicates the presence of a vortex inside the triangle at position $(x_{v},y_{v})$. 
Divisions by zero  in Eq. \eqref{eq-circ-tri} are avoided in programs by first checking the value of the wave function on the vertices of the triangle. This enables the algorithm to also capture vortices passing exactly through a vertex of the mesh.

\subsection{Vortex radius estimation}\label{sec-2D-fitReg}

The difficulty of this step is the separation of the vortex density from the background (Thomas-Fermi) density. We use the decomposition:
\begin{equation}\label{eq-uvortex}
\rho = \rho_{v} \rho_{b}
\end{equation}
with $\rho_{b}$ is the background density corresponding to the shape of the condensate and $\rho_{v}$ the vortex density  varying between 0 and 1. The computation of {$\rho_{b}$} is done using two methods.
\begin{itemize}
	\item If the parameters defining the trapping potential are known, we use again the Thomas-Fermi approximation, as in Sect. \ref{sec-vortex-localization}. For this case ({\em quadratic-minus-quartic} potential) we approximate $\rho_b=\rho_\TF$ using \eqref{eq-TF-detail}.

	\item If the Thomas-Fermi expression cannot be reconstructed analytically, we assume the general formula:
	\begin{equation}
	\label{eq-fit-TF}
	\rho_b = \rho_0^\fit - \frac{1}{2} \left(a_x^\fit x^2 + a_y^\fit y^2 + a_4^\fit (x^2 + y^2)^2\right),
	\end{equation}
	and use a least-square procedure to find coefficients $C=\left\{\rho_0^\fit, a_x^\fit, a_y^\fit, a_4^\fit \right\}$ minimizing the functional
	\begin{equation}
	J(C) = \frac{1}{2} || \rho(x,y) - \rho_b(C,x,y)||_{Th_\trunc}^2.
	\label{eq-JC}
	\end{equation}
\end{itemize}
Note that the minimization is performed on the domain ${Th_\trunc}$ representing the condensate after removing the regions corresponding to vortex domains (where the density has a sharp decrease to zero). 
A minimization on the entire domain $Th$ resulted in a poor approximation of the Thomas-Fermi density, when tested against analytical forms. 
Several algorithms defining ${Th_\trunc}$ were tested and finally the most reliable and fast algorithm is based on removing individual vortex domains which are disks of radius $0.1 d_{\min}$, where $d_{\min}$ is the minimum distance between vortex centers. The minimization of the functional \eqref{eq-JC} is easily computed using the interface of \ff with the  \texttt{Ipopt} library based on the interior point minimisation method  \citep{Ipopt-line-search-2005}.

The obtained background density $\rho_b$ is plotted  in Fig. \ref{fig-2D-quartic}(b). It is now possible to isolate vortex density and fit it with a Gaussian, following
\begin{equation}
\rho_b - \rho= \rho_b (1-\rho_v)=A \exp\left[-\frac{1}{2}\frac{(x-{x_0})^2}{{r_x}^2}-\frac{1}{2}\frac{(y-{y_0})^2}{{r_y}^2}\right].
\label{eq-fit-Gauss}
\end{equation}
The amplitude $A$ is used to define the vortex contrast (as in experiments, \cite{BEC-physV-2001-codd}),
since  $A/\rho_b(x_0,y_0)$ is the ratio between the "missing"
column density at vortex center  and the corresponding TF
value (see also \cite{dan-2005}). Only vortices with a contrast greater than 0.3 are
considered to compute core radii $r_v=(r_x+r_y)/2$. Note that we used for the vortex fit \eqref{eq-fit-Gauss} an anisotropic Gaussian. In practice $r_x \approx r_y$. As the vortex center position has been already  computed, the fit is simplified by setting $x_0 = x_v, y_0 = y_v$. When fitting the vortex center position during tests, we obtained the expected behaviour: $x_0 \approx x_v$ and $y_0 \approx y_v$. The fit following Eq. \eqref{eq-fit-Gauss} is computed on individual vortex domains which are disks of radius  $0.45 d_{\min}$ centered on vortex center, as shown in Fig. \ref{fig-2D-quartic}(c).  The \texttt{Ipopt} library is again used for the least-square fit, which ensures fast and reliable results. 

The program also computes the characteristics of  the vortex lattice (vortex arrangement and vortex lattice parameter).  A mesh is built using  vortex points as vertices, as shown in Fig. \ref{fig-2D-quartic}(d), on the top of the original density plot. This representation is useful to better distinguish the arrangement of vortices in different types of vortex lattices (circular, triangular,  squared) than commonly used plots of contours of density. For the case of dense Abrikosov lattices (as shown in Sect. \ref{sec-Ex-Abrikosov}), we select vortices with 6 neighbours and compute the  vortex lattice parameter as the average distance to these 6 neighbours.

\section{Vortex identification in 3D}\label{sec-3D}

\begin{figure}
	\centering
	\begin{subfigure}[b]{0.48\textwidth}
		\centering
		\includegraphics[width=\textwidth]{\figpath/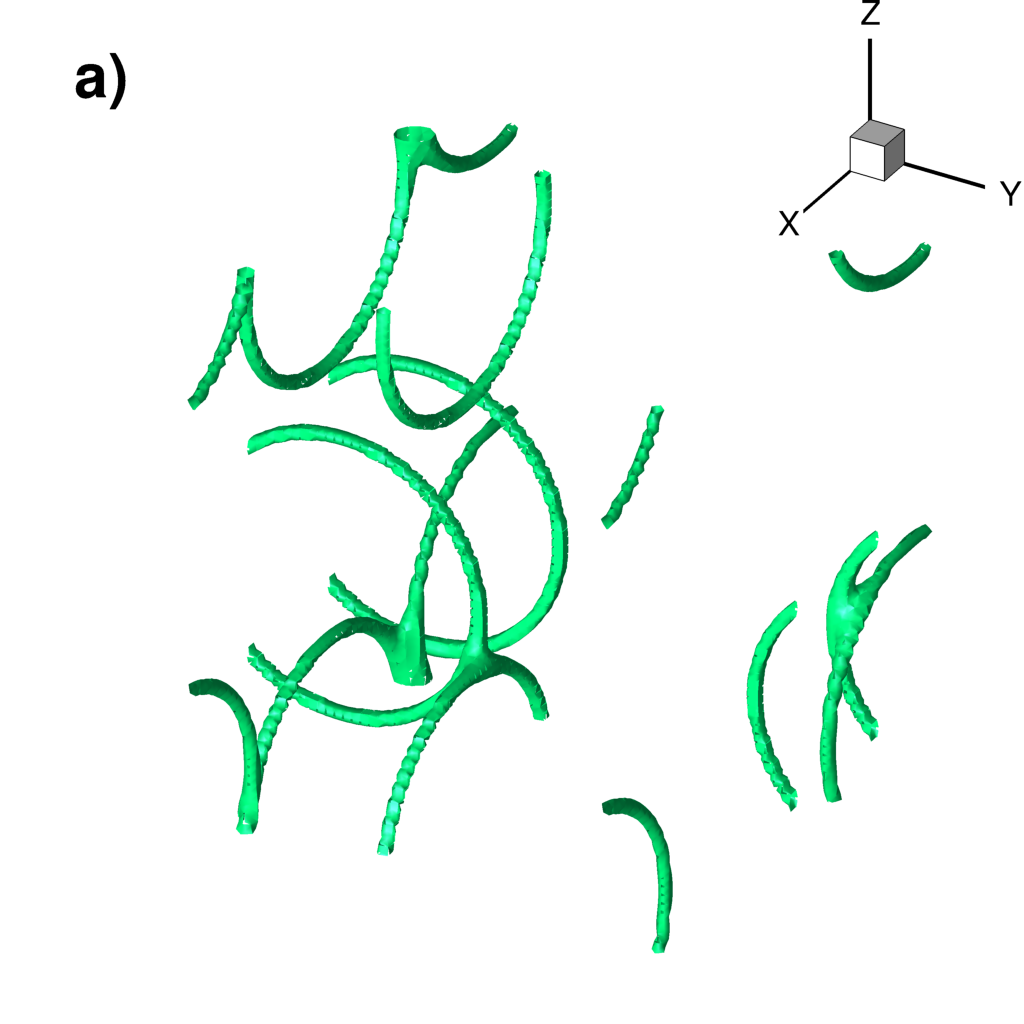}
		\label{fig:gps_a}
	\end{subfigure}
	\hfill
	\begin{subfigure}[b]{0.48\textwidth}
		\centering
		\includegraphics[width=\textwidth]{\figpath/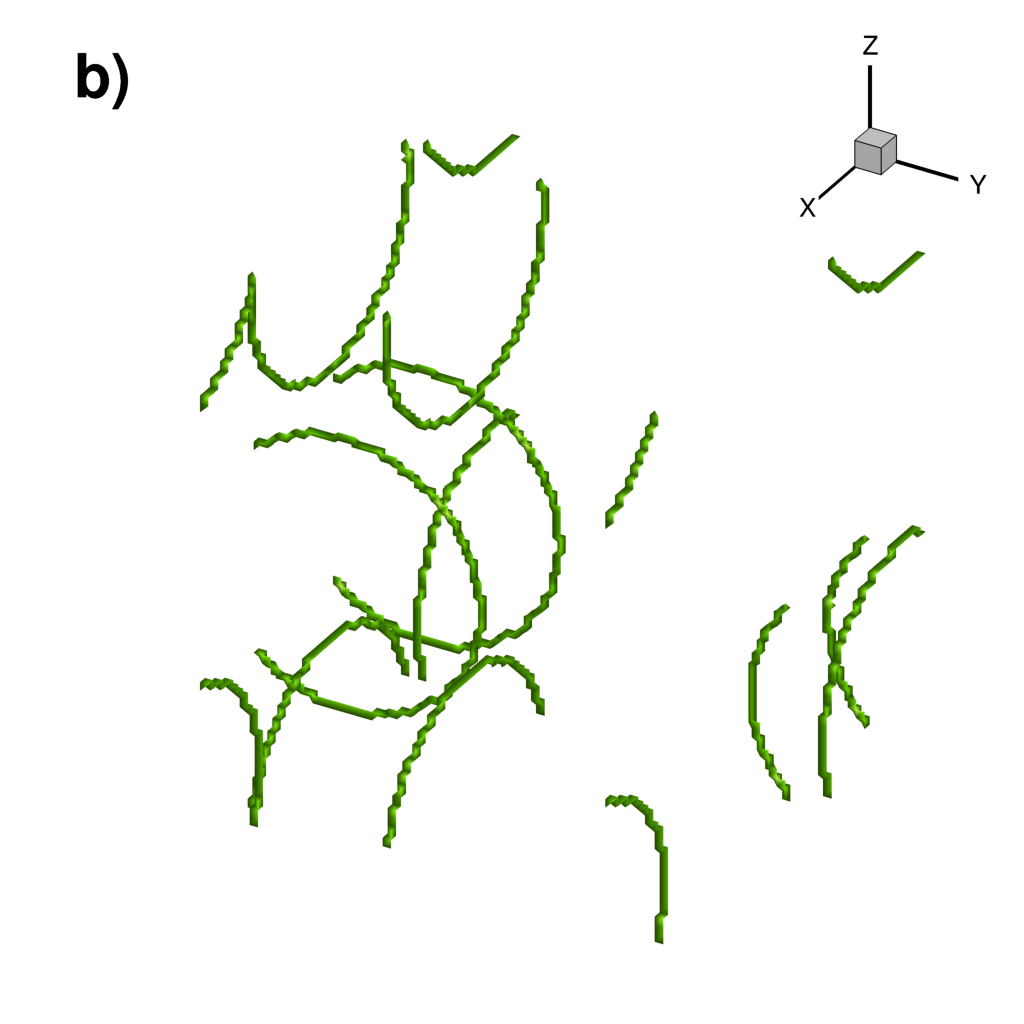}
		\label{fig:gps_b}
	\end{subfigure}
	\begin{subfigure}[b]{0.48\textwidth}
		\centering
		\includegraphics[width=\textwidth]{\figpath/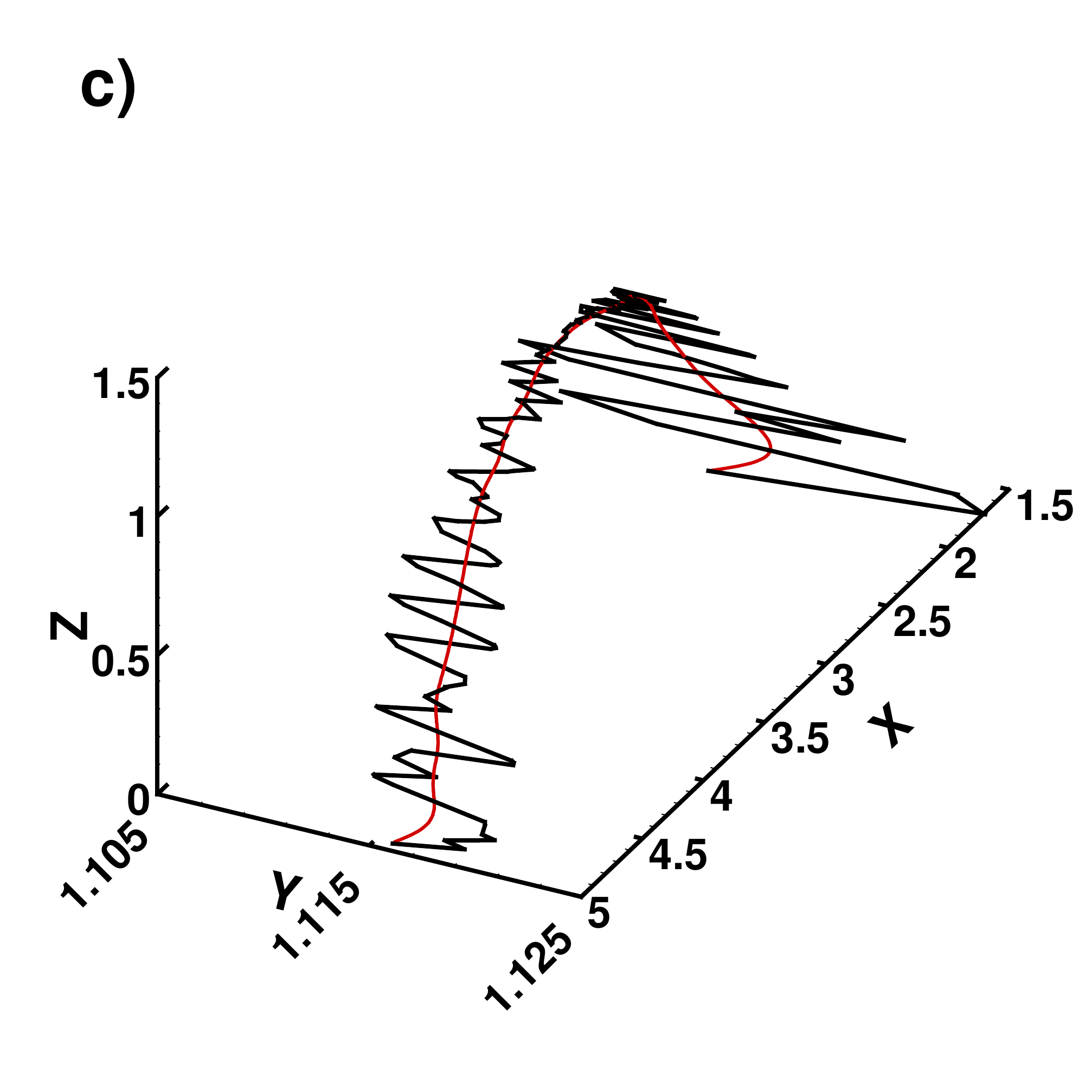}
		\label{fig:gps_c}
	\end{subfigure}
	\hfill
	\begin{subfigure}[b]{0.48\textwidth}
		\centering
		\includegraphics[width=\textwidth]{\figpath/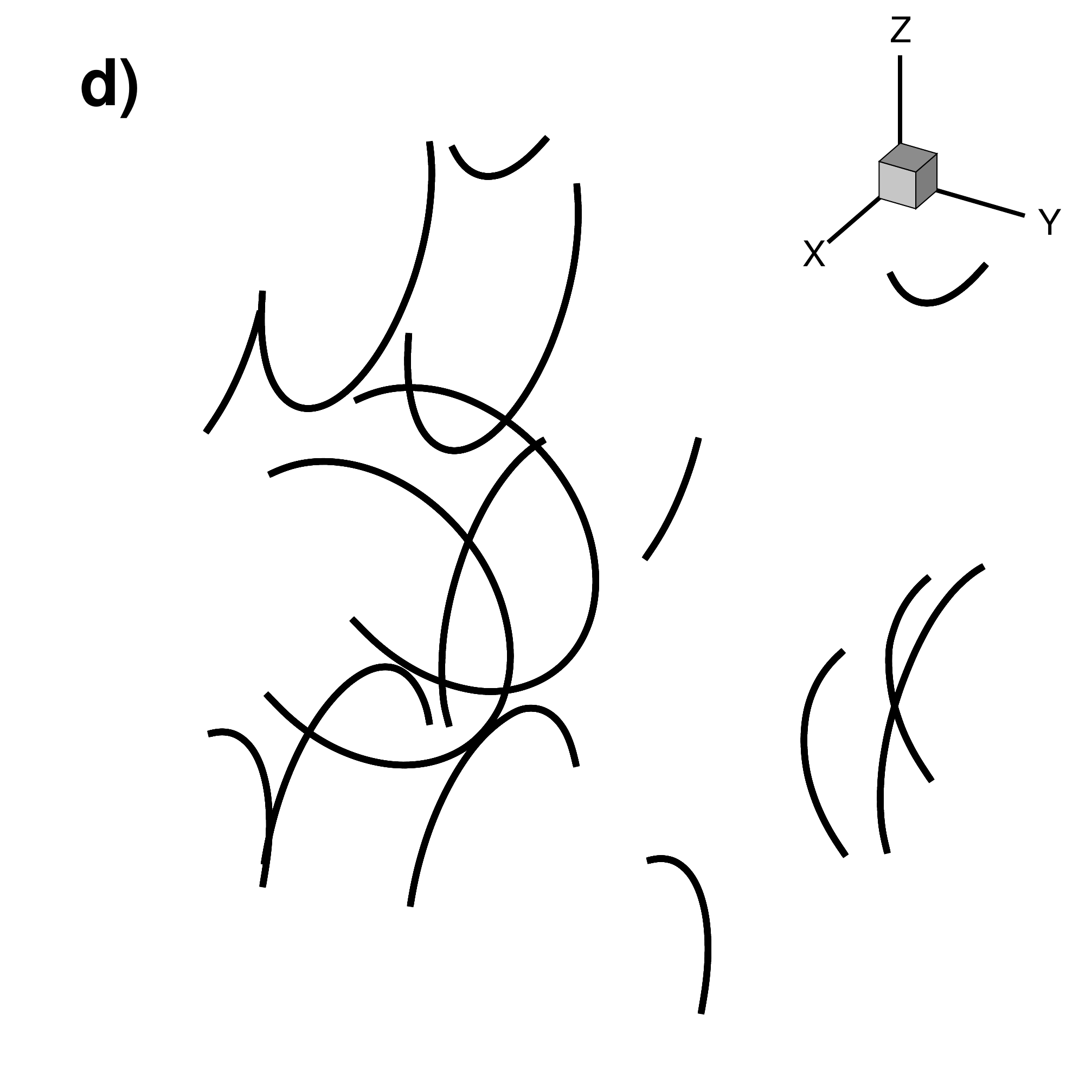}
		\label{fig:gps_d}
	\end{subfigure}
	\caption{Illustration of the vortex identification algorithm in 3D. Test case with vortex rings in superfluid helium \citep{dan-2021-CPC-QUTE} simulated inside a cubic domain (spectral method). a) Isosurface of  low density $\rho$, b) tetrahedrons crossed by vortex lines, c) example of extracted vortex line (black) and smoothed vortex line (red), d) final vortex lines reconstructed in the full domain.}
	\label{fig-3D-gps1}
\end{figure}

Vortex identification  in 3D combines the localization algorithm described in Sect. \ref{sec-vortex-localization} with the graph-based method suggested by \cite{liu2020vectorizing}. The wave function is represented as a P1 finite element function on mesh elements which are now tetrahedrons. Vortex lines are described by a series of points identified as the intersection of vortex lines with faces of a mesh element. The algorithm is illustrated in Fig. \ref{fig-3D-gps1} using numerical data corresponding to a  superfluid helium field with vortex rings \citep{dan-2021-CPC-QUTE}. A low-density isosurface of the initial data is presented in Fig. \ref{fig-3D-gps1}(a). An important difference when studying vortex lines instead of points is the presence of reconnections (vortex line intersection) and closed loops.

The algorithm used for the localization of vortices in 2D is directly adapted to 3D by applying Eqs. (\ref{eq-zero-tri}), (\ref{eq-zero-point}) and (\ref{eq-circ-tri}) on the faces of each tetrahedron. The result is the detection of faces crossed by vortex lines. The position of the zero density point inside those faces is computed and stored. Tetrahedrons crossed by  vortex lines are plotted in Fig. \ref{fig-3D-gps1}(b). Following the method suggested by \cite{liu2020vectorizing}, a graph is then built by linking the points belonging to a same mesh element and using the fact that a given point is inside two different tetrahedrons. The connectivity $c$ of the graph allows an easy separation of vortex lines through the following algorithm:
\begin{enumerate}
	\item Choose a point of connectivity $c=1$ and follow the graph while keeping an ordered list of visited points. Stop when reaching a point where $c=1$ (end of a line) or $c>2$ (reconnection point). Remove all visited points from the graph except if the end point is a reconnection point; in this case remove the link to the penultimate point and actualize connectivity. Repeat this step until there are no  points with $c=1$ left. The ordered lists of points represent  vortex lines.
	\item The only remaining lines are now loops. Start by removing the ones that reconnect through the same process as before, excepting that we start and stop at points where $c>2$.
	\item Only points with $c=2$ remain. Starting from one such point and stopping when reaching the same point ends the identification of a vortex loop. All lines have been extracted from the graph when all points have been extracted.
\end{enumerate}

As the points are constrained to be on faces of tetrahedrons, the vortex line is not smooth, as illustrated in Fig. \ref{fig-3D-gps1}(c). Two possibilities exist to smooth the vortex line. For a precise study of  vortex lines, it is necessary to locate more accurately its points through the interpolation of the wave function and a local minimization of the density to identify its zero. This local minimization can be done using a Newton method  \citep{villois2016vortex} or by combining a pseudo-vorticity method and a gradient descent minimization \citep{liu2020vectorizing}. The drawback of these methods is that they are time-consuming and directly related to the spectral representation of the solution. We privileged a second possibility to smooth the vortex line, using a simple five-point moving average along the line. This methodology proved to be very fast and sufficiently precise to identify 3D vortex lines. 
Since this is a low-cost methodology, the results thus obtained could be eventually used as initial (guess) data in  more complex  procedures, as those suggested  by \cite{villois2016vortex,liu2020vectorizing},  to identify vortex lines with spectral accuracy.

For a line $l$, with points $l_i,\ 1\leq i \leq n$, we compute new points $m_i$, as follows:
\begin{equation}
\label{eq-smooth-line}
m_i = \frac{0.8}{5}\left( l_{i-2} + l_{i-1} + l_i + l_{i+1} + l_{i+2}\right) + 0.2 l_i.
\end{equation}
The ends of the line are kept fixed and a three point window is used for the second and penultimate points. The process is repeated, with the number of iterations being an adjustable parameter. In practice, $10$ iterations are usually enough. To obtain a smoother result, it is possible to use a linear interpolation on  line segments to increase the number of points before applying the moving average. Figure \ref{fig-3D-gps1}(c) shows a comparison between the initial line and the smoothed result for one of the vortex lines (located at the bottom left of the initial solution). The $y$ axis of the figure is enlarged for the clarity of the visualisation. All lines identified in the original  field are plotted in Fig. \ref{fig-3D-gps1}(d).

An optional quantity that can be computed is the curvature $\gamma$ of the vortex line. At a given point $m_i$, we estimate $\gamma(m_i) = {\alpha}/{\Delta s}$, where $\Delta s$ is the arc-length variation between points $m_{i-1}$ and $m_{i+1}$ and $\alpha$ is the angle between line segments linking $m_{i-1}$ to $m_i$ and $m_i$ to $m_{i+1}$.

\clearpage
\section{Vortex identification in experimental images}\label{sec-2D-images}

\begin{figure}
	\centering
	\begin{subfigure}[b]{0.32\textwidth}
		\centering
		\includegraphics[width=\textwidth]{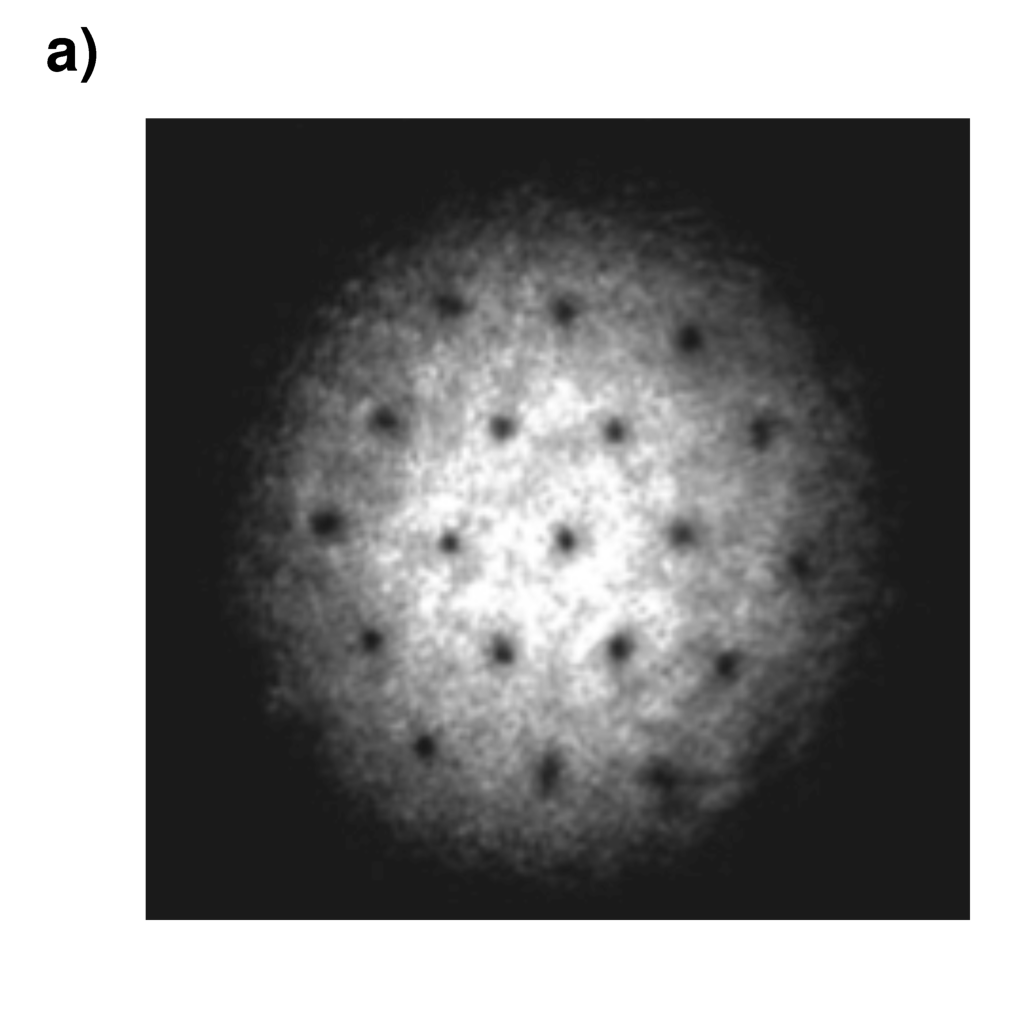}
		\label{fig:small_latt_a}
	\end{subfigure}
	\hfill
	\begin{subfigure}[b]{0.32\textwidth}
		\centering
		\includegraphics[width=\textwidth]{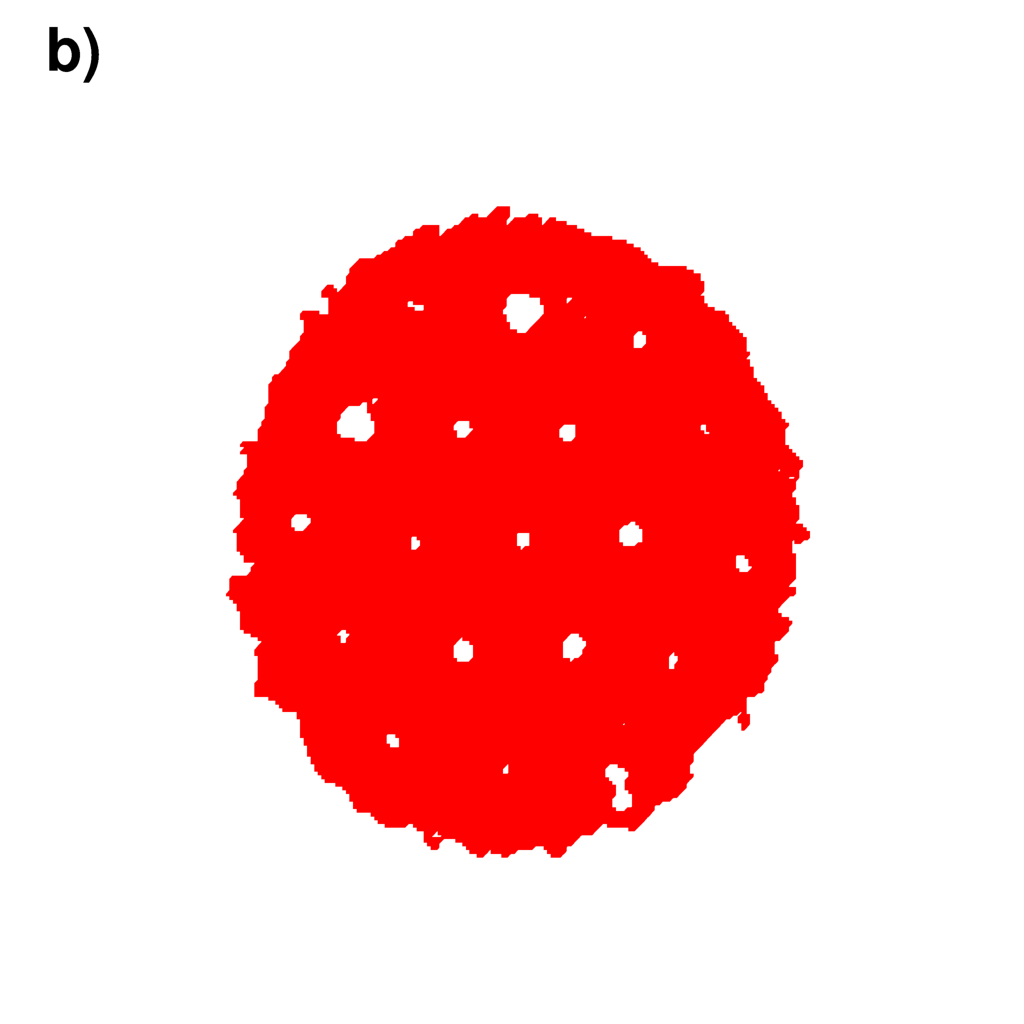}
		\label{fig:small_latt_b}
	\end{subfigure}
	\begin{subfigure}[b]{0.32\textwidth}
		\centering
		\includegraphics[width=\textwidth]{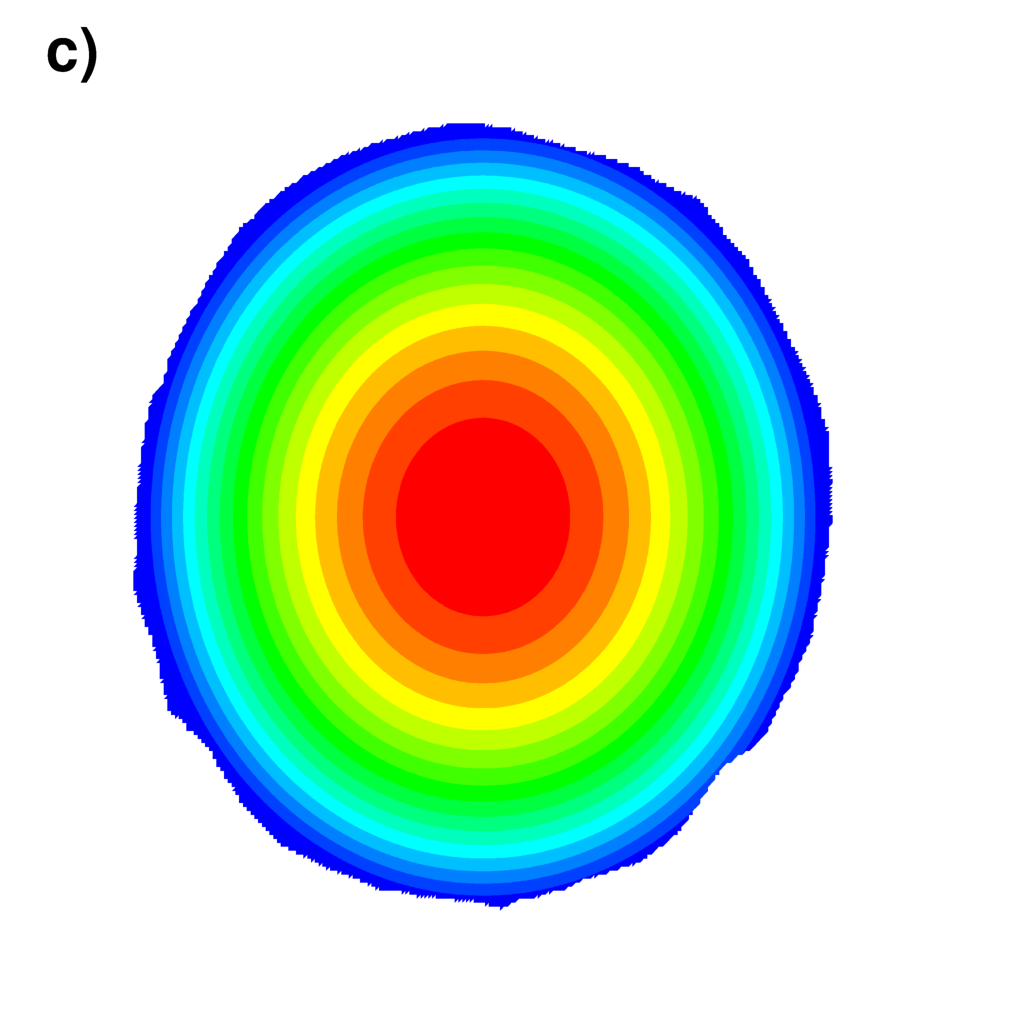}
		\label{fig:small_latt_c}
	\end{subfigure}
	\hfill
	\begin{subfigure}[b]{0.48\textwidth}
		\centering
		\includegraphics[width=\textwidth]{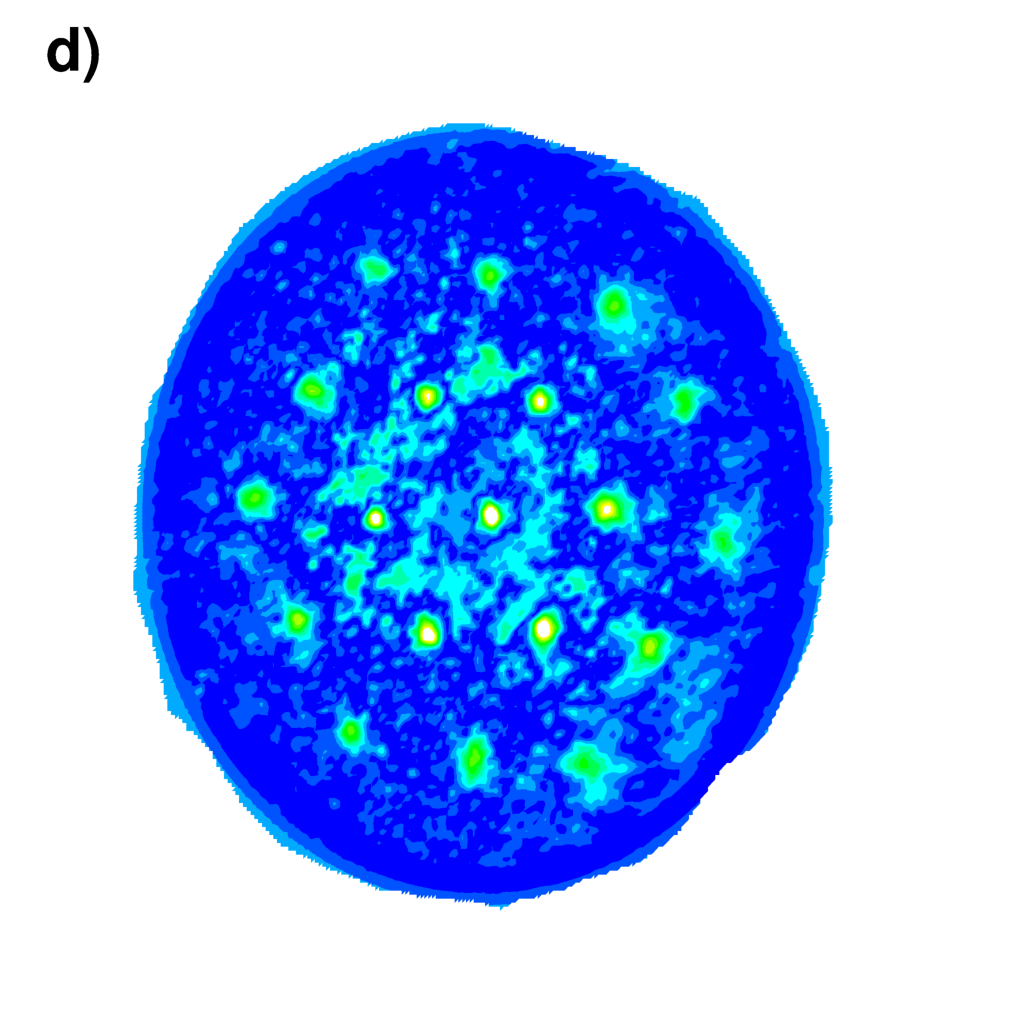}
		\label{fig:small_latt_d}
	\end{subfigure}
	\hfill
	\begin{subfigure}[b]{0.48\textwidth}
		\centering
		\includegraphics[width=\textwidth]{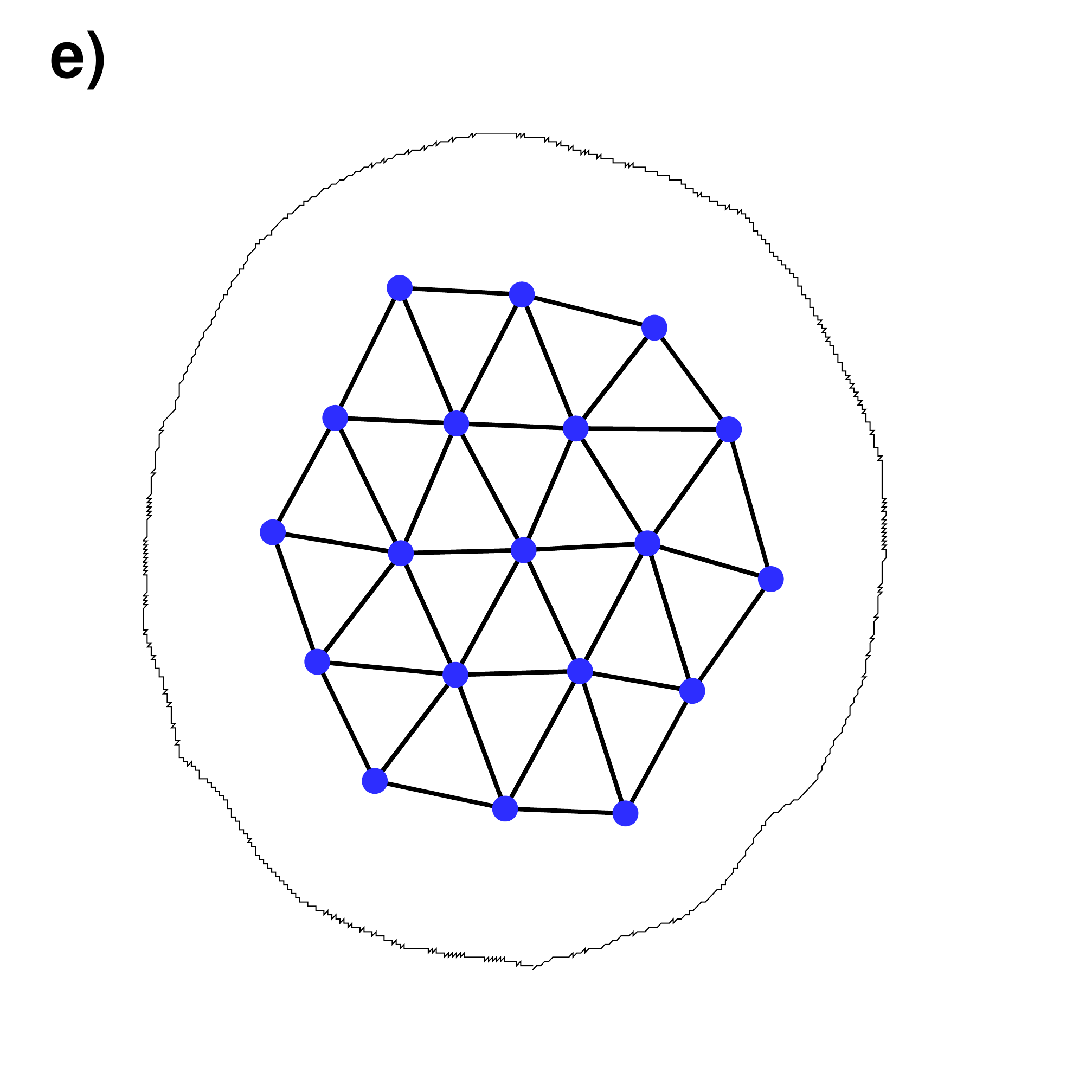}
		\label{fig:small_latt_e}
	\end{subfigure}
	\caption{Illustration of  vortex identification steps in experimental images. Case of a vortex lattice in a slowly rotating BEC \citep{BEC-physV-2001-codd}. a) Experimental image of atomic density $\rho$, b) region separated using a density contour $\rho_{iso}$, c) fitted background (Thomas-Fermi) density $\rho_\TF$, d) vortex density $\rho_{v}$, e) mesh of identified vortices (blue points) and condensate border.}
	\label{fig-2Dimg-small_latt}
\end{figure}

The toolbox distributed with this paper can be also used to analyse experimental images of BEC with vortices. Such images are obtained by switching-off the magnetic trap and imaging the absorption of a resonant laser beam along the vertical axis. The contrast in experimental pictures thus represents the atomic density integrated along the $z$-axis and vortices are visible as very dark regular spots (see Fig. \ref{fig-2Dimg-small_latt}).  Since the information concerning the phase of the wave function is not available, the algorithms presented in Sect. \ref{sec-2D-data} have to be adapted  to analyse experimental images. 

We illustrate the adapted algorithm by extracting vortex characteristics from an experimental image presented by  \cite{BEC-physV-2001-codd}.  It corresponds to a slowly rotating atomic BEC in an harmonic trap.  The original image, shown in Fig. \ref{fig-2Dimg-small_latt}(a), is transformed in the \texttt{pgm} format and then loaded in \ff as a rectangular array of grey values between 0 (dark) and 1 (white). A triangular finite-element mesh is then constructed from the rectangular grid and the values of grey (representing the density contrast) are represented as a P1 (piece-wise linear) function on this mesh.

The first step of the algorithm consists in separating  vortex regions from the rest of the domain. We define a new variable $\rho_{iso} = \mathbb{1}_{\rho > c_{iso}} - \mathbb{1}_{\rho \leq c_{iso}}$, where $\mathbb{1}$ is the characteristic (indicator) function and $c_{iso}$ is an adjustable parameter. 
If it is necessary to remove noisy artefacts on the border of the condensate, $\rho_{iso}$ can be smoothed using algorithms inspired from  image processing techniques. We implemented a smoothing algorithm that balances between the accuracy of the fit of the contour and the preservation of the length of the contour, but any other image smoothing algorithm can be tested. Figure \ref{fig-2Dimg-small_latt}(b) shows that $\rho_{iso}$ separates the mesh into three subdomains that can be identified as the condensate, the vortices domain (holes) and the exterior of the condensate. If some vortices are not correctly separated from the background contrast, the value $c_{iso}$ can be adjusted by trials and the above algorithm repeated.


For each vortex region, the vortex center $(x_v,y_v)$ is computed as the minimum of the density in the region. Following this point, all the information needed to apply the algorithms presented in Sect. \ref{sec-2D-fitReg} are available. 
\begin{itemize}
	\item A mesh ${Th_\trunc}$ is created by removing circles of radius $0.1 d_{min}$ around each vortex point, where $d_{min}$ is the minimum distance between vortices.
	\item The background density $\rho_b$ is then estimated following Eqs. \eqref{eq-fit-TF} and \eqref{eq-JC}. This density is plotted in Fig. \ref{fig-2Dimg-small_latt}(c).
	\item The vortex density $\rho_{v}$, presented in Fig. \ref{fig-2Dimg-small_latt}(d), is fitted with the Gaussian \eqref{eq-fit-Gauss} to estimate vortex radius $r_v$. To obtain a better estimation of the position of the vortex center, $x_0$ and $y_0$ are also parameters of the fit (which was not the case for the fit of 2D numerical data). The mesh of identified vortices is plotted in Fig. \ref{fig-2Dimg-small_latt}(e).
\end{itemize}

\section{Examples and benchmarks provided with the toolbox}\label{sec-example}

Several other examples and benchmarks used to validate our numerical codes are provided with the toolbox. To reduce the size of the files distributed with the 3D code, solutions have been truncated to remove tetrahedrons where $\rho > \rho_{threshold}$, with $\rho_{threshold}$ being adapted to each simulation to ensure that a high density area is still surrounding each vortex. Table \ref{tab-time} presents the computational (CPU) time and the number of mesh elements for the different test cases. The first test cases in each category correspond to  examples used to explain algorithms in Sects. \ref{sec-2D-data}, \ref{sec-3D} and \ref{sec-2D-images}.\enlargethispage{5\baselineskip}
\begin{table}[!h]
	\begin{tabular}{clcll}
		\multicolumn{1}{l}{}     & Test case                     & Figure & CPU time (s) & Number of triangles \\ \hline
		\multirow{2}{*}{2D data} & Quartic - quadratic potential & \ref{fig-2D-quartic}& 2.5 (14.7)              & 32258                            \\
		& Dense Abrikosov lattice      & \ref{fig-2D-Abrikosov}       & 22.6 (79.6)              & 130050                           \\ \hline
		\multirow{6}{*}{3D data}  & Vortex rings       & \ref{fig-3D-gps1}           & 0.29           & 16574 (4368)                     \\
		& Bose-Einstein condensate  & \ref{fig-3D-BEC}    & 1.4              & 267963 (84562)                   \\
		& Vortex knot                 & \ref{fig-3D-knot}  & 0.5            & 95004 (3479)                      \\
		& $k^{-3}$ Kelvin wave    & \ref{fig-3D-KW}      & 1.08             & 273733 (912)                      \\
		& $k^{-6}$ Kelvin wave    & \ref{fig-3D-KW}      & 1.09              & 273673 (834)                     \\
		& $128^3$ quantum turbulence  & \ref{fig-3D-QT}  & 3.9              & 101189 (37756)                   \\
		\multicolumn{1}{l}{}     & $256^3$ quantum turbulence  & \ref{fig-3D-QT}  & 774.7              & 99488250 (345541)                \\ \hline
		experimental                  & Small Abrikosov lattice   & \ref{fig-2Dimg-small_latt}    & 13.82            & 91574                            \\
		images      & Dense Abrikosov lattice    & \ref{fig-2D-lattice}   & 27.66              & 114720                          
	\end{tabular}
	\caption{Computational (CPU) time and number of mesh elements for different test cases. For 2D data, the indicated time corresponds to the computation of $\rho_{b}$ using numerical parameters. The computational time when using a  fit to estimate $\rho_{b}$ is indicated in parenthesis. For 3D cases, the number of tetrahedrons crossed by a vortex line is indicated in parenthesis.}
	\label{tab-time}
\end{table}

\clearpage

\subsection{Dense 2D Abrikosov vortex lattice}\label{sec-Ex-Abrikosov}

\begin{figure}[!h]
	\centering
	\begin{subfigure}[b]{0.45\textwidth}
		\centering
		\includegraphics[width=\textwidth]{\figpath/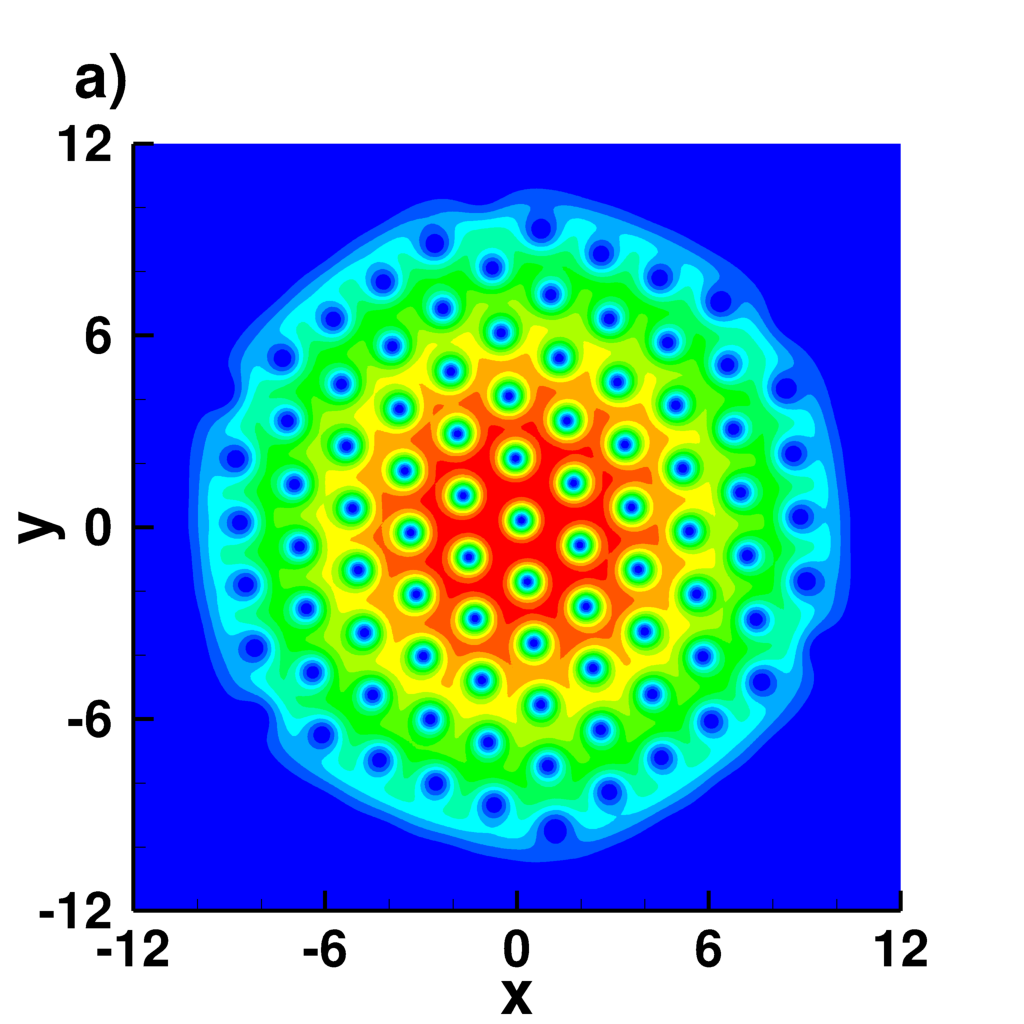}
		\label{fig:Abrikosov_a}
	\end{subfigure}
	\hfill
	\begin{subfigure}[b]{0.45\textwidth}
		\centering
		\includegraphics[width=\textwidth]{\figpath/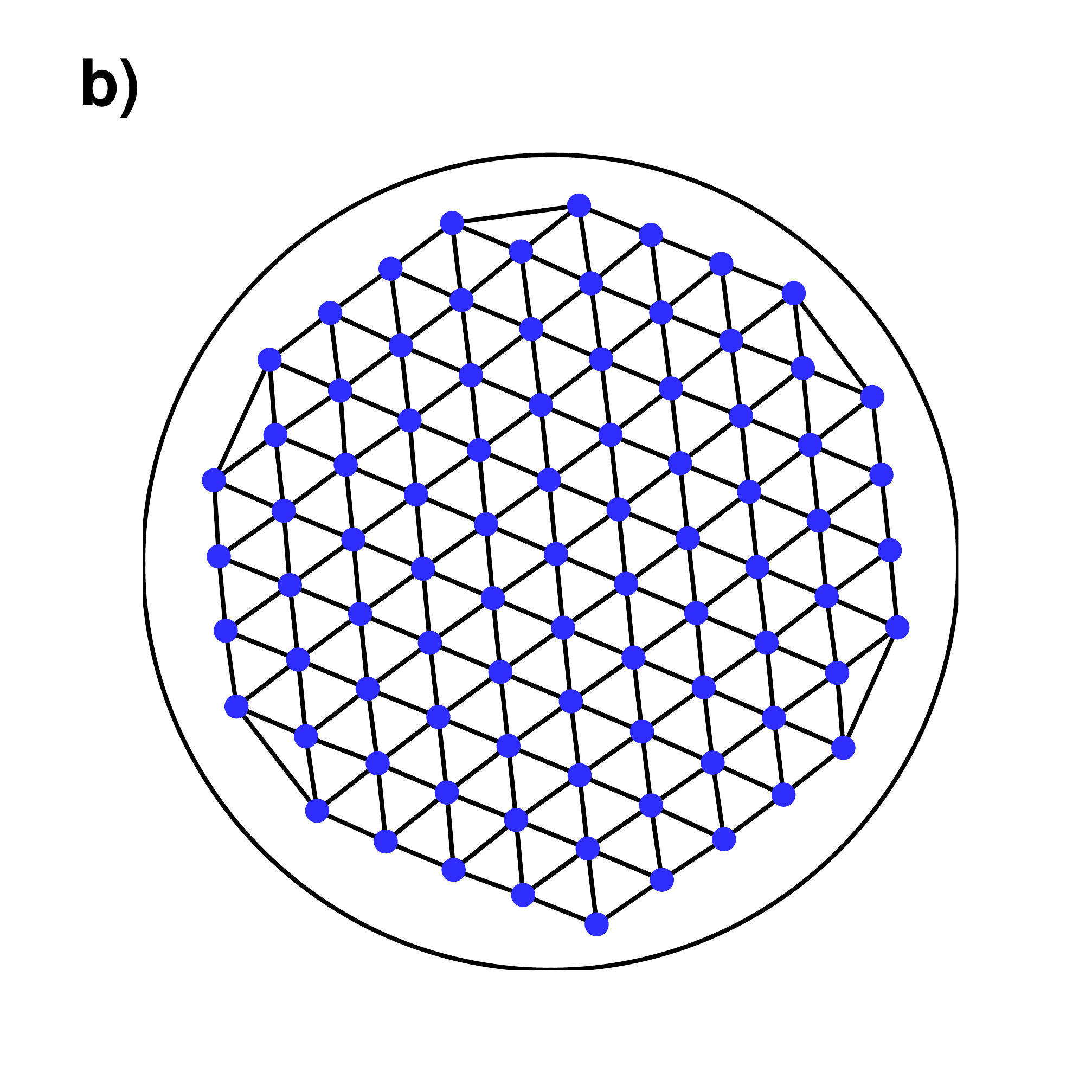}
		\label{fig:Abrikosov_b}
	\end{subfigure}
	\begin{subfigure}[b]{0.45\textwidth}
		\centering
		\includegraphics[width=\textwidth]{\figpath/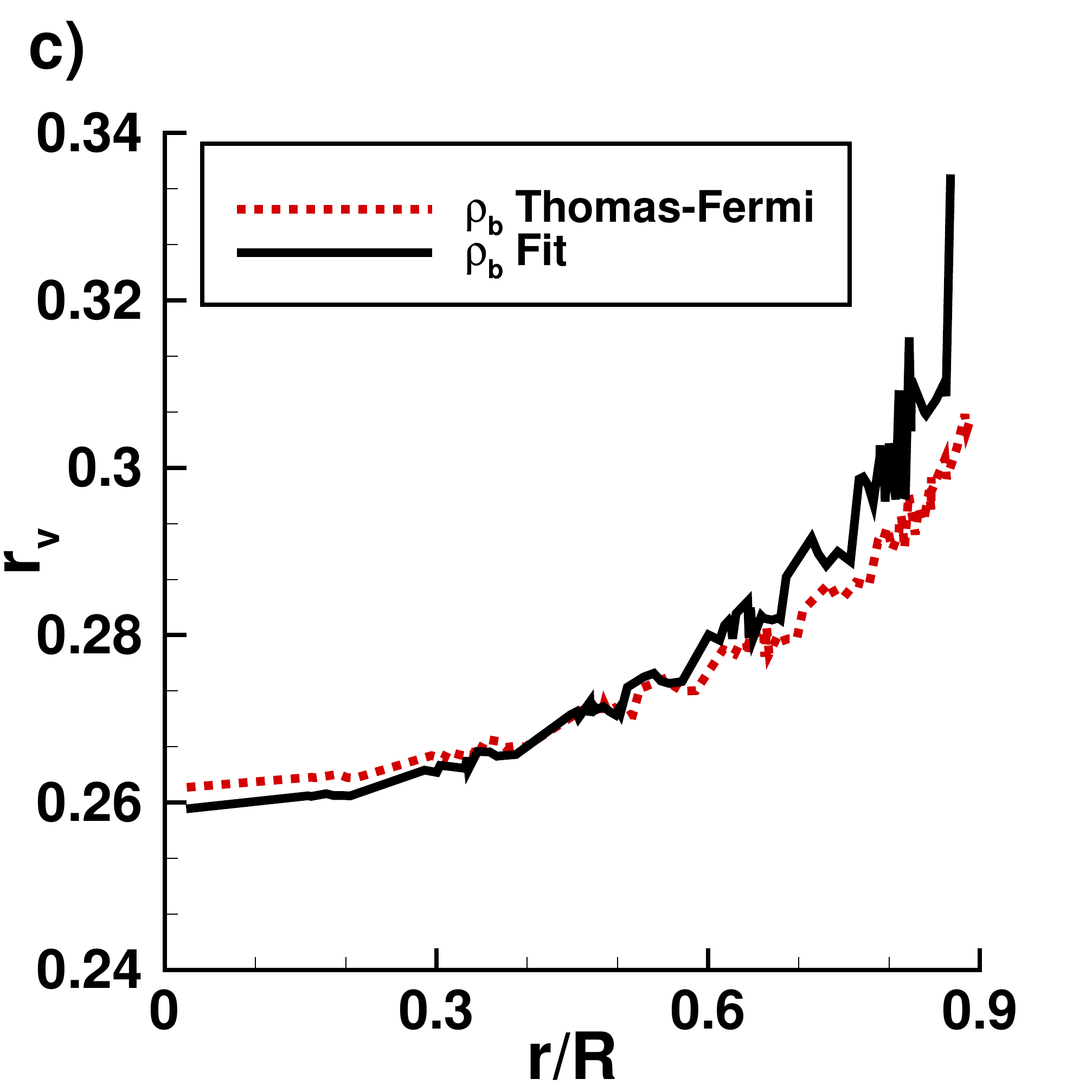}
		\label{fig:Abrikosov_c}
	\end{subfigure}
	\hfill
	\begin{subfigure}[b]{0.45\textwidth}
		\centering
		\includegraphics[width=\textwidth]{\figpath/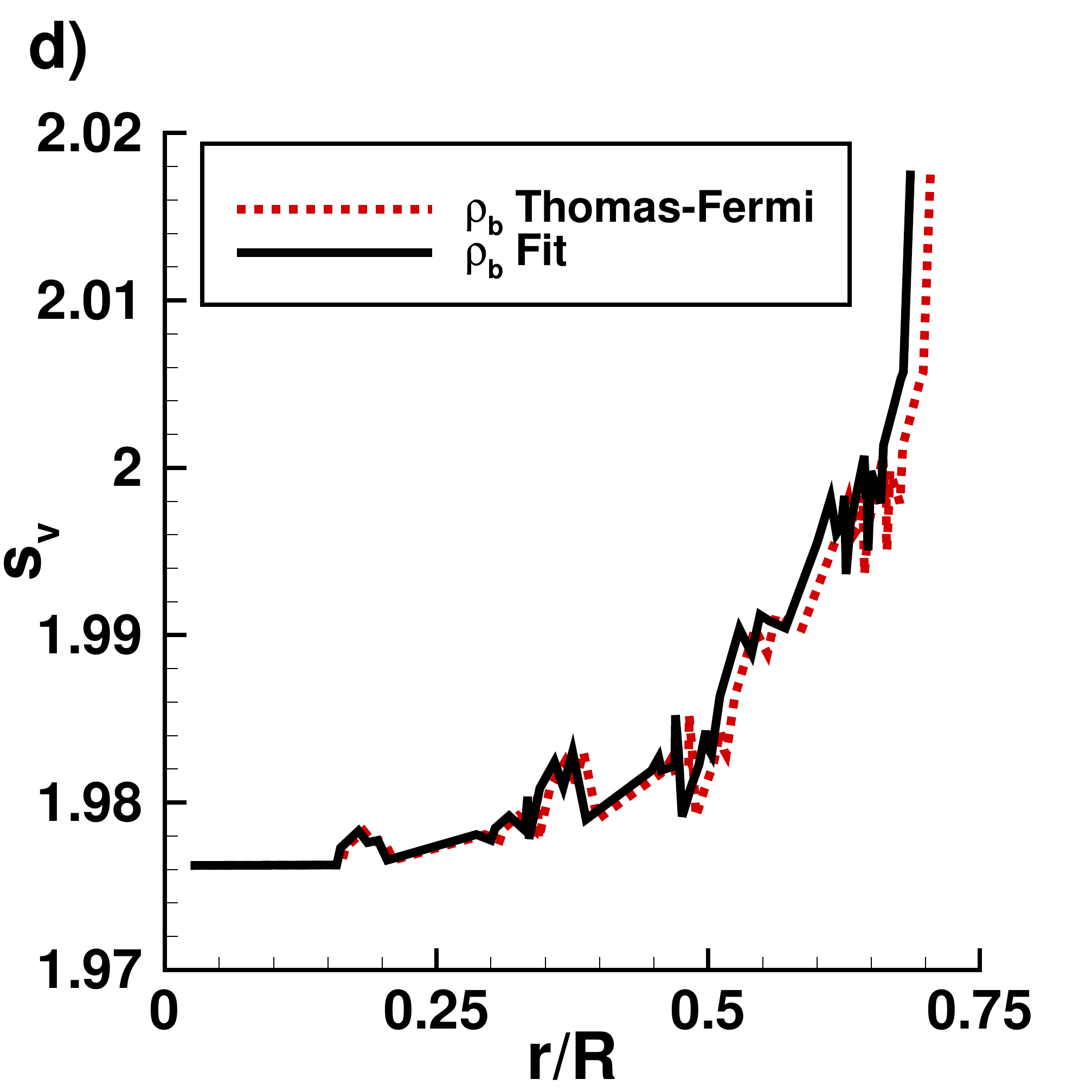}
		\label{fig:Abrikosov_d}
	\end{subfigure}
	\caption{Illustration of the vortex identification in 2D for a dense Abrikosov lattice in fast rotating BEC. Wave function field obtained with a spectral GP code \citep{Parnaudeau_2015}. a) Initial density $\rho$, b) mesh of identified vortices, c) vortex radius $r_v$ as a function of the normalized distance from the vortex to the center of the condensate ${r}/{R_\TF}$, d) inter-vortex distance $s_v$.}
	\label{fig-2D-Abrikosov}
\end{figure}

This case corresponds to a 2D Abrikosov lattice in a fast rotating BEC. The wave function field was obtained using a spectral solver \citep{Parnaudeau_2015} for the scaled Gross-Pitaevskii equation \eqref{eq-scal-GP} with parameters: $C_g = 1000$, $C_\Omega = 0.95$, $a_x = a_y = 1$. The background density $\rho_{b}$ is obtained by computing the Thomas-Fermi density directly from simulation parameters. The initial data is shown in Fig. \ref{fig-2D-Abrikosov}(a) and the identified vortices are plotted in Fig. \ref{fig-2D-Abrikosov}(b). Characteristics of the Abrikosov lattice are plotted in next panels:  the vortex radius $r_v$ as a function of the distance to the condensate center ${r}/{R_\TF}$ in Fig. \ref{fig-2D-Abrikosov}(c) and the inter-vortex distance $s_v$ in Fig. \ref{fig-2D-Abrikosov}(d). 

\pagebreak

\subsection{Vortices in 3D Bose-Einstein condensates}\label{sec-Ex-bec}

\begin{figure}[!h]
	\centering
	\begin{subfigure}[b]{0.48\textwidth}
		\centering
		\includegraphics[width=\textwidth]{\figpath/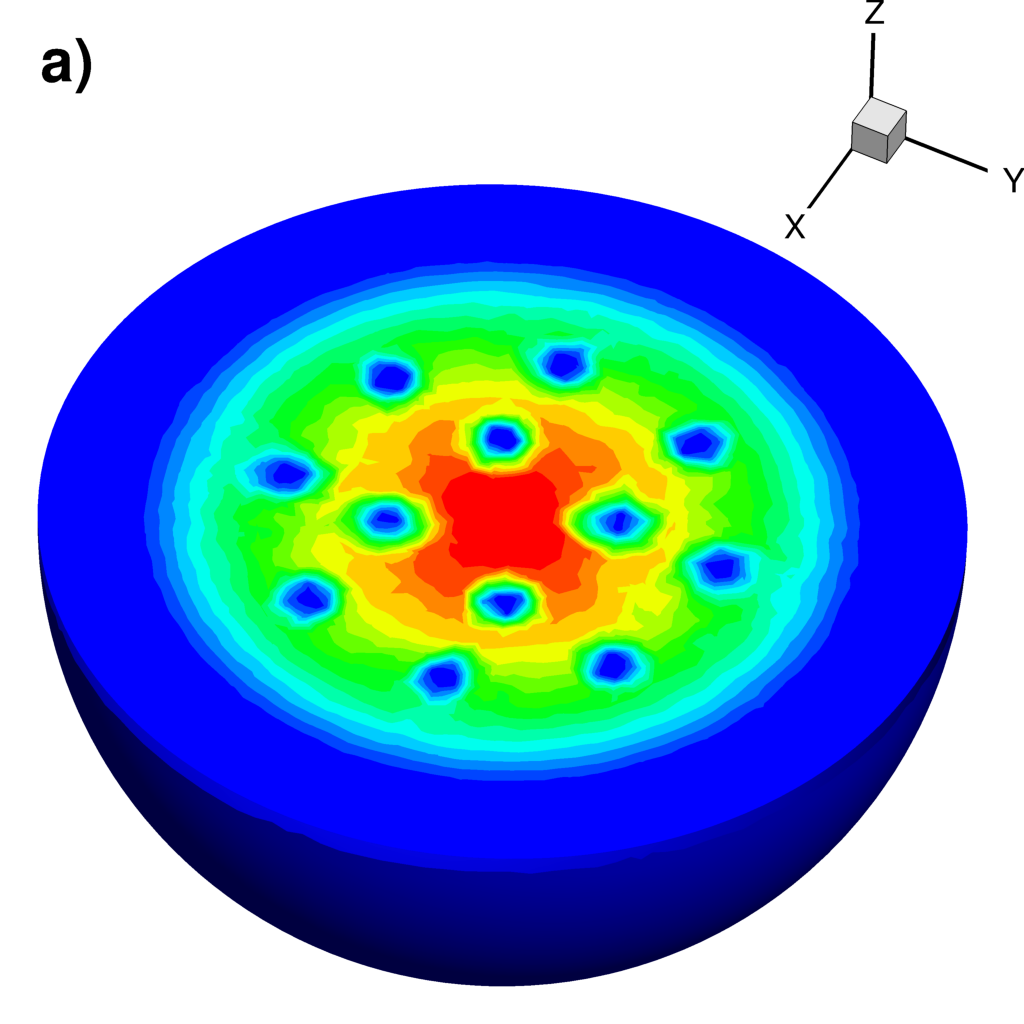}
		\label{fig:bec_a}
	\end{subfigure}
	\hfill
	\begin{subfigure}[b]{0.48\textwidth}
		\centering
		\includegraphics[width=\textwidth]{\figpath/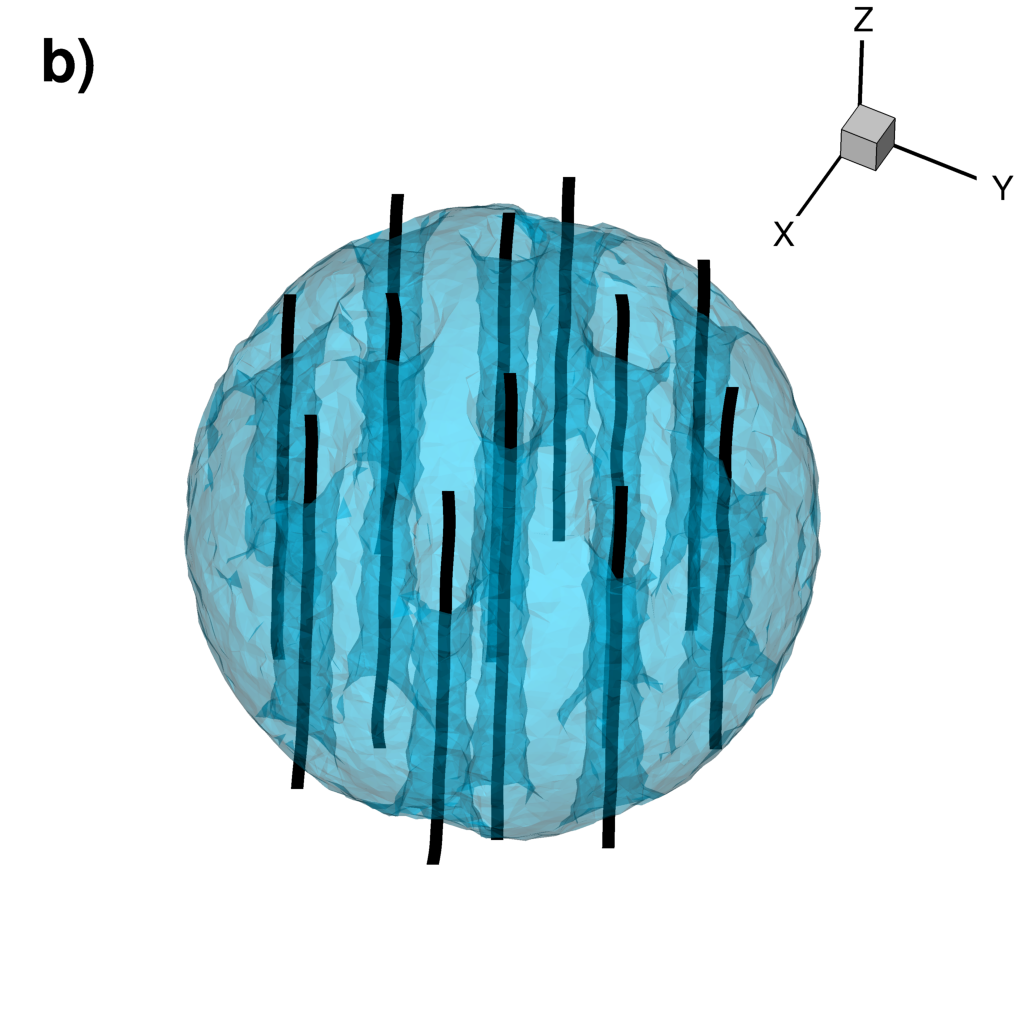}
		\label{fig:bec_b}
	\end{subfigure}
	\caption{Illustration of the vortex line identification in 3D for a rotating BEC. Data obtained on an unstructured mesh using a finite-element solver \citep{dan-2016-CPC}. a) Cut of the initial density $\rho$, b) extracted vortex lines.}
	\label{fig-3D-BEC}
\end{figure}

To test the toolbox on an unstructured mesh, we simulated a 3D BEC with the GP finite-element toolbox distributed by  \cite{dan-2016-CPC}. A cut of the initial data with 12 vortices is shown in Fig. \ref{fig-3D-BEC}(a). The extracted lines are plotted in Fig. \ref{fig-3D-BEC}(b).

\subsection{3D vortex knot}\label{sec-Ex-VRknot}

\begin{figure}[!h]
	\centering
	\begin{subfigure}[b]{0.48\textwidth}
		\centering
		\includegraphics[width=\textwidth]{\figpath/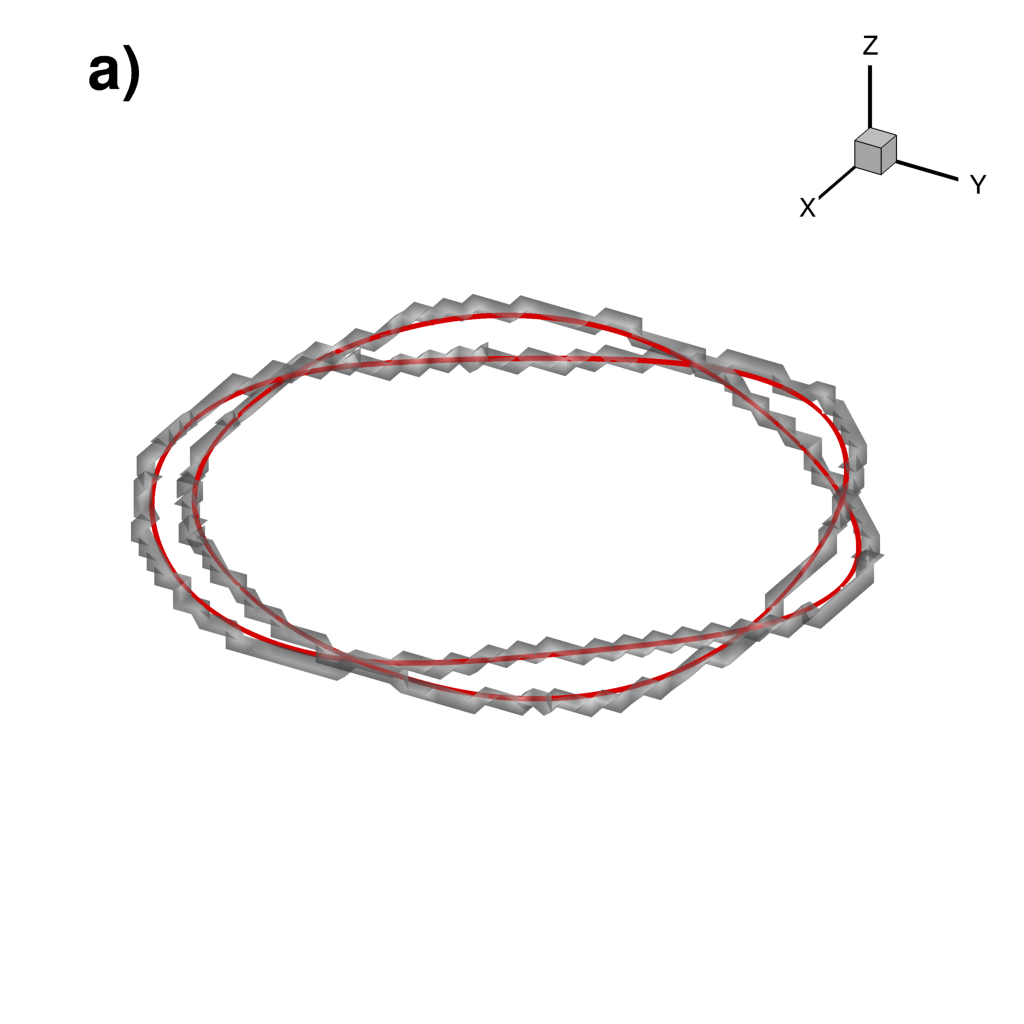}
		\label{fig:VRknot_a}
	\end{subfigure}
	\hfill
	\begin{subfigure}[b]{0.48\textwidth}
		\centering
		\includegraphics[width=\textwidth]{\figpath/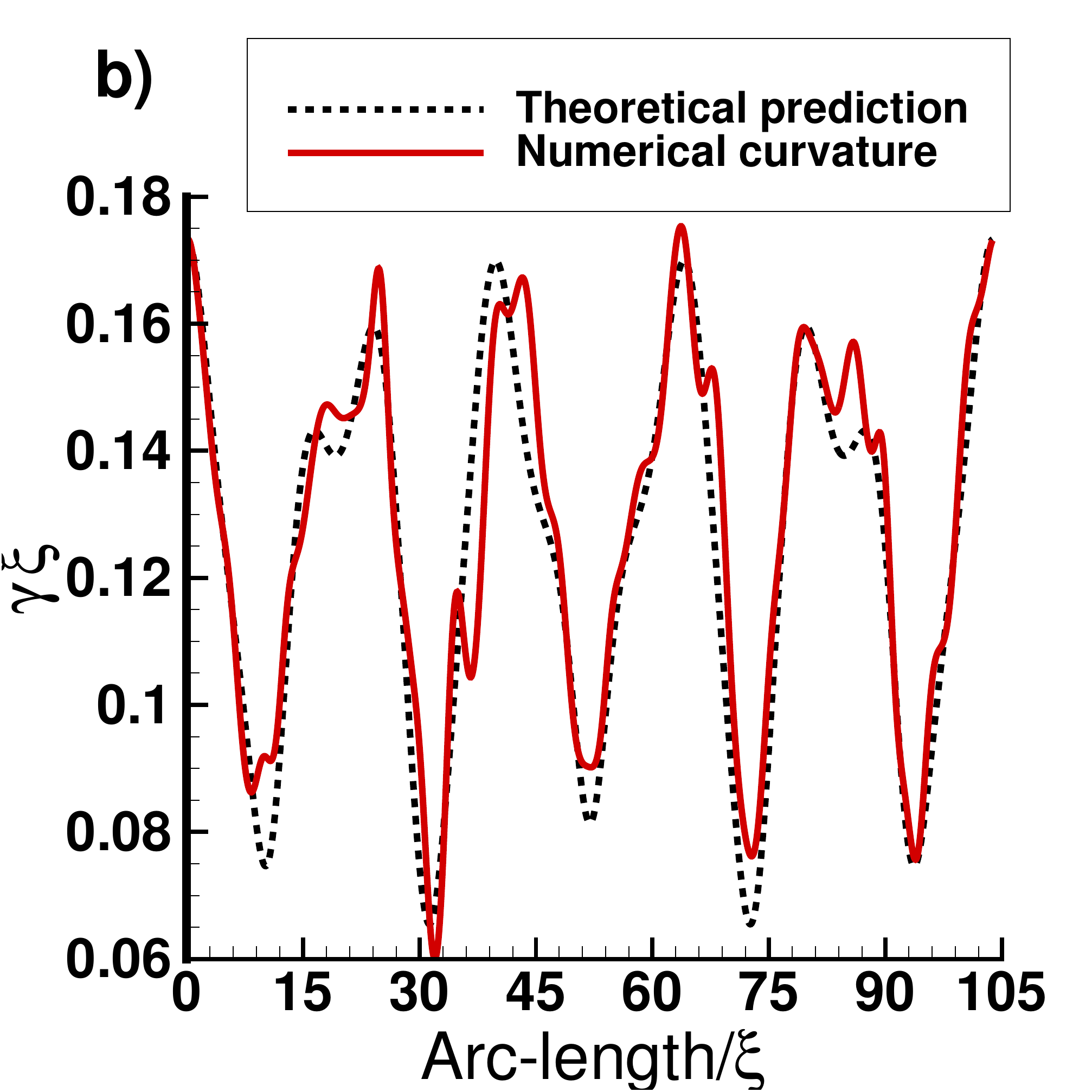}
		\label{fig:VRknot_b}
	\end{subfigure}
	\caption{Illustration of the vortex line identification in 3D for a vortex knot. Test case suggested by  \cite{villois2016vortex}. a) Vortex line (in red) inside the mesh elements identified of circulation $\kappa=1$, b) comparison of the vortex line curvature with analytical values.}
	\label{fig-3D-knot}
\end{figure}

For a more quantitative assessment of the results provided by the toolbox, we tested the programs for a case with known analytical description of  the vortex line. This is the vortex knot benchmark presented by \cite{villois2016vortex}. The vortex line is defined as $s(\sigma) = (x_p^{-1}(s_x(\sigma)), y_p^{-1}(s_y(\sigma)), z_p^{-1}(s_z(\sigma)))$ with:
\begin{align}\label{eq-knot}
s_x(\sigma) &= \left[R_0 + R_1\cos(q\sigma) \right] \cos(p\sigma),\\
s_y(\sigma) &= \left[R_0 + R_1\cos(q\sigma) \right] \sin(p\sigma),\\
s_z(\sigma) &= R_1\sin(q\sigma).
\end{align}
Functions $x_p$, $y_p$ and $z_p$ are introduced during the computation of the wave function to ensure periodic boundary conditions:
\begin{align}
x_p &= - \frac{L}{\pi}\cos\left(\frac{\pi}{L}x\right),\\
y_p &= -\frac{L}{\pi}\cos\left(\frac{\pi}{L}y\right),\\
z_p &= - \frac{L}{2\pi}\cos\left(\frac{2\pi}{L}z\right).
\end{align}
A comparison between the vortex  line described analytically by Eq. \eqref{eq-knot} and the result obtained with the toolbox  is shown in Fig. \ref{fig-3D-knot}(a). The wave function was computed with a $128^3$ grid, in a cubic domain of side $2\pi$. The knot parameters are: $R_0 = 8\xi$, $r_1 = \xi/2$, $p=2$ and $q=5$ with $\xi = 0.5\Delta x$. The curvature $\gamma$ of the vortex line was computed as an additional test and compared in Fig. \ref{fig-3D-knot}(b) to analytical values. A fairly good approximation is observed. To reach the precision obtained by \cite{villois2016vortex}, a higher resolution and a better smoothing method are needed (with the drawback of considerably increasing the computational time). The script used to create the initial wave function for this case is included in the toolbox.

\subsection{Kelvin waves}\label{sec-Ex-Kwave}

Another validation against analytical forms of vortex lines considered Kelvin waves with a $k^{-3}$ and $k^{-6}$ energy spectrum. Oscillations of the vortex line were generated following the method suggested by \cite{proment2013interaction}. A Pad{\'e} approximation of the vortex profile  \citep{berloff2004pade} was used in each $xy$ plane. The vortex center in each plane was placed along a line $(X(z),Y(z),z)$ defined by:
\begin{equation}
X(z) + iY(z) = \mathcal{F}^{-1}\left( \sqrt{{n_{th}}(k)}e^{i\phi(k)}\right),
\end{equation}
where ${n_{th}}(k)$ is the imposed spectrum (here $k^{-3}$ or $k^{-6}$) and $\phi(k)$ is randomly distributed in $[0,2\pi[$. $\mathcal{F}^{-1}$ denotes the inverse Fourier transform.

\begin{figure}[!h]
	\centering
	\begin{subfigure}[b]{0.48\textwidth}
		\centering
		\includegraphics[width=\textwidth]{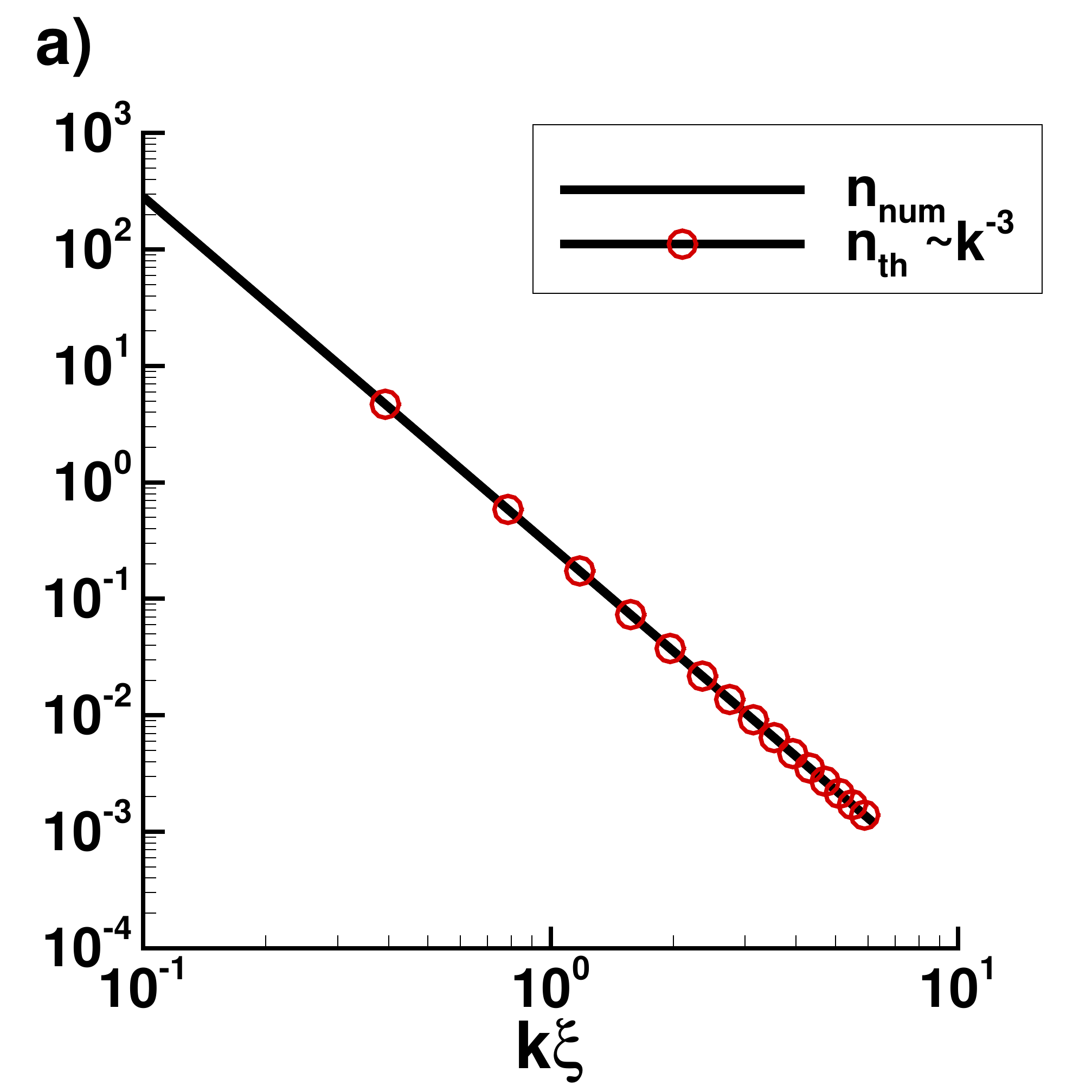}
		\label{fig:KWave_a}
	\end{subfigure}
	\hfill
	\begin{subfigure}[b]{0.48\textwidth}
		\centering
		\includegraphics[width=\textwidth]{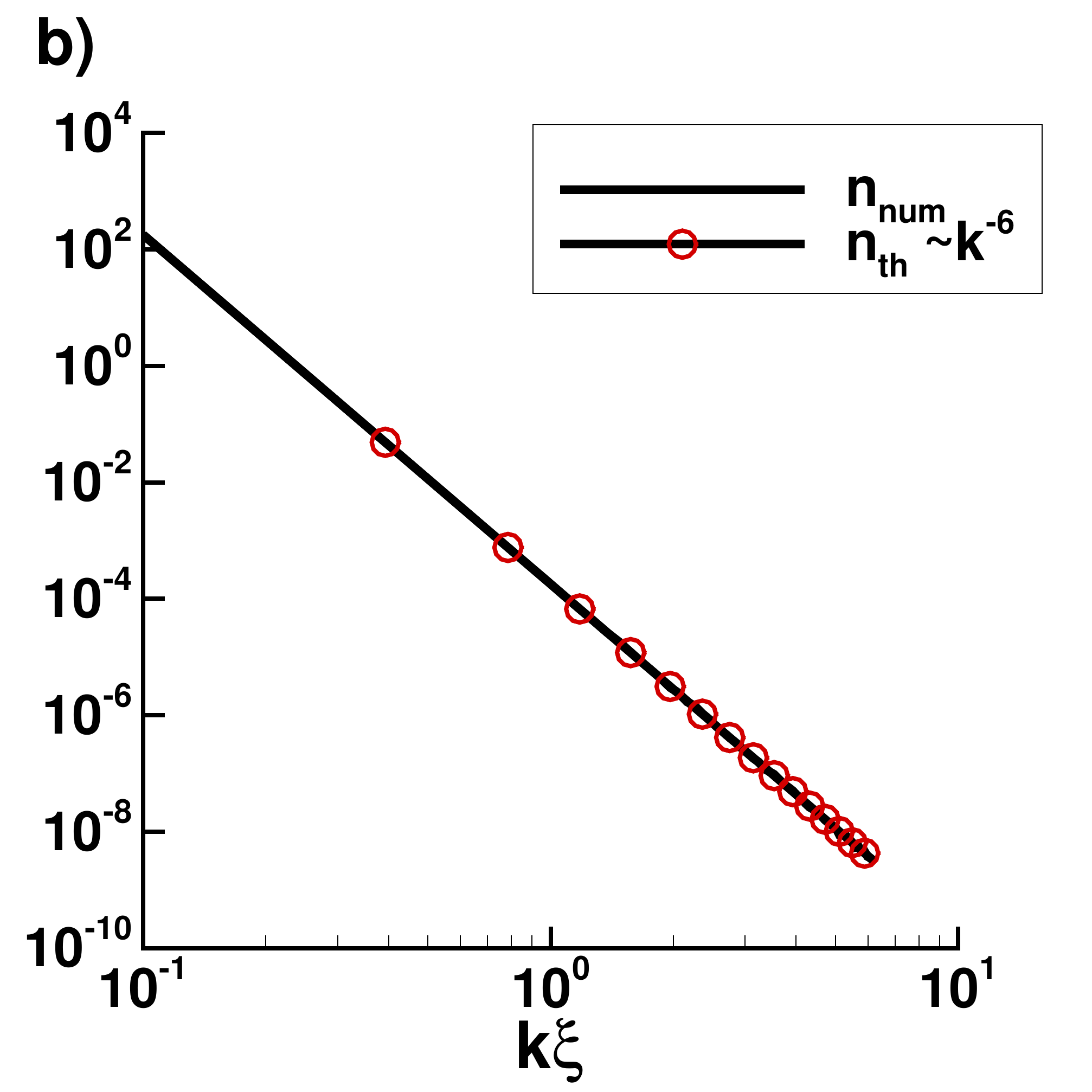}
		\label{fig:KWave_b}
	\end{subfigure}
	\begin{subfigure}[b]{0.48\textwidth}
		\centering
		\includegraphics[width=\textwidth]{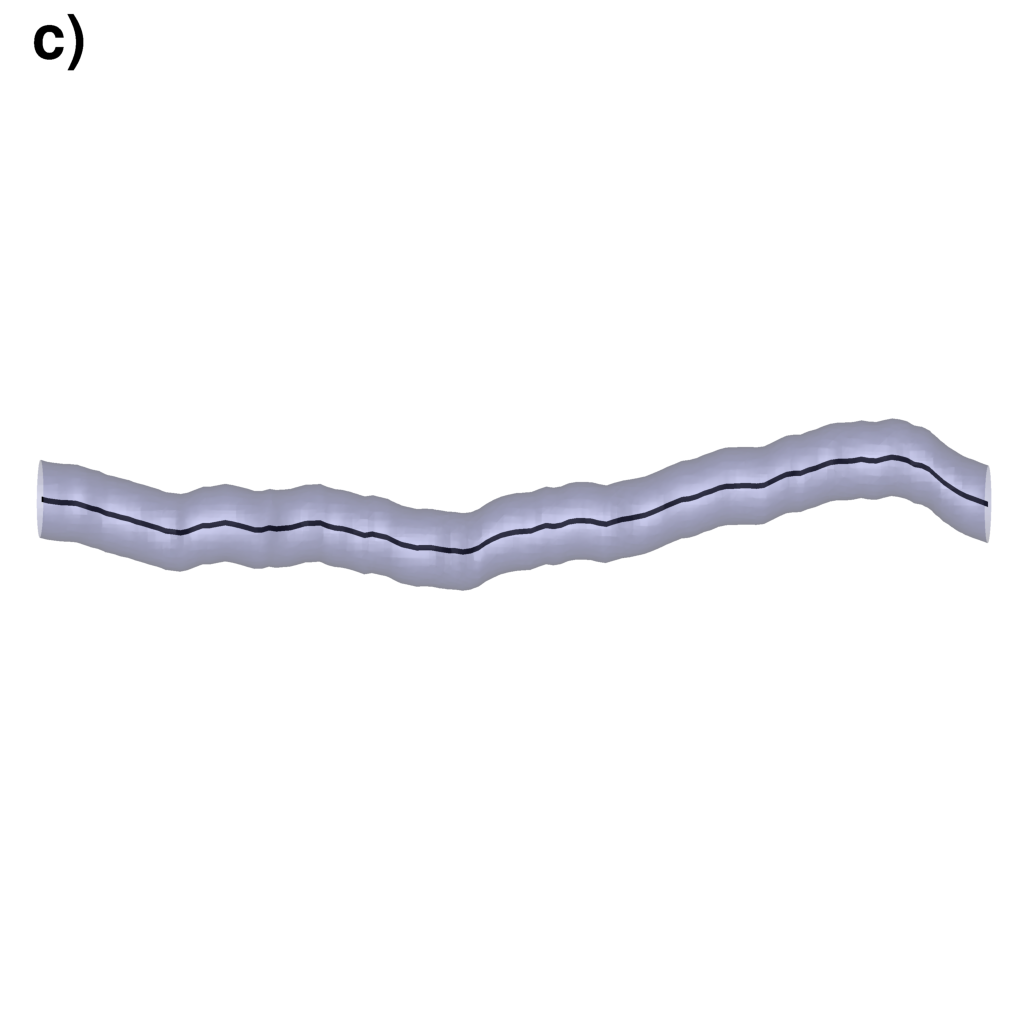}
		\label{fig:KWave_c}
	\end{subfigure}
	\hfill
	\begin{subfigure}[b]{0.48\textwidth}
		\centering
		\includegraphics[width=\textwidth]{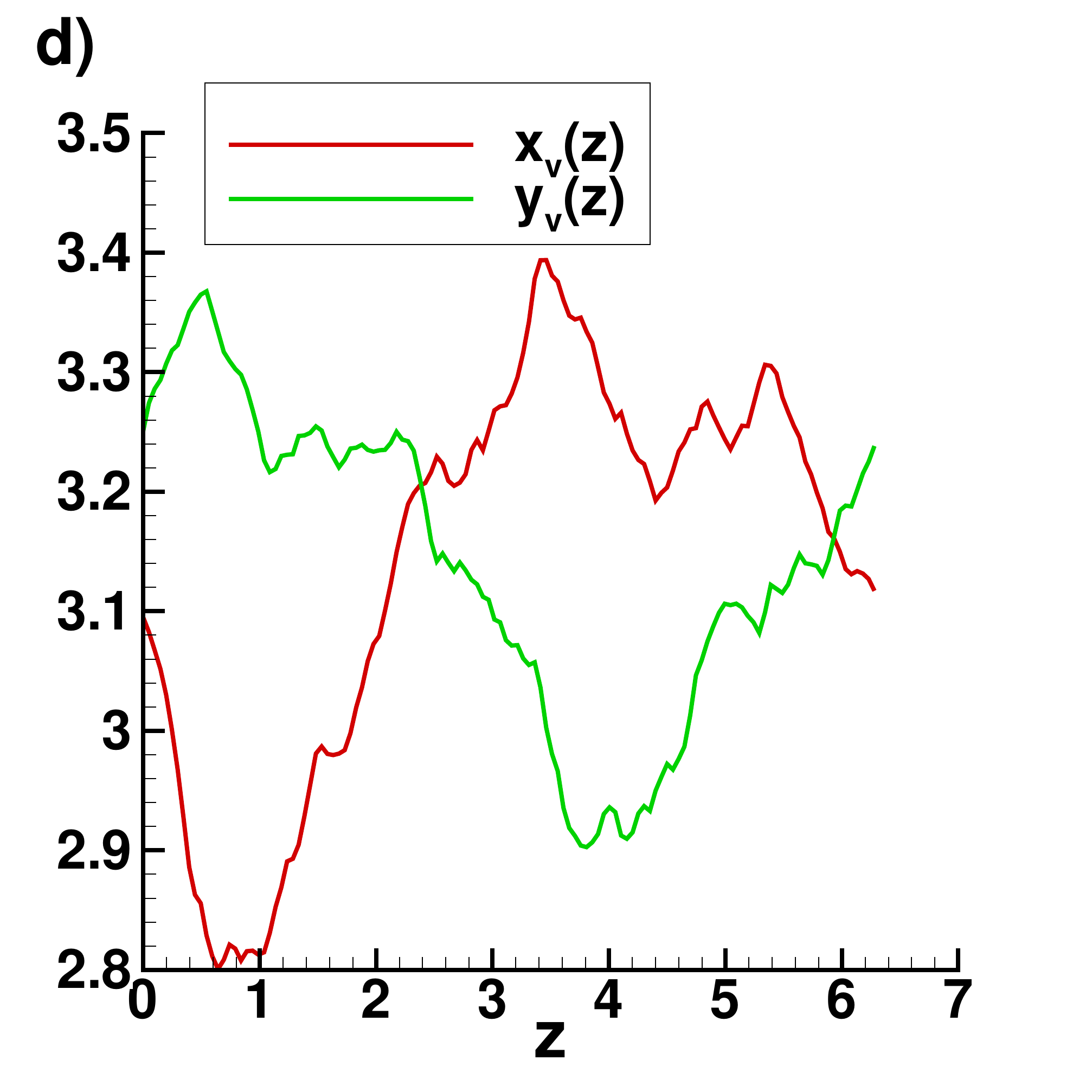}
		\label{fig:KWave_d}
	\end{subfigure}
	\caption{Kelvin wave spectrum for a vortex line scaling as $k^{-3}$ (a) and $k^{-6}$ (b). Test case suggested by \cite{proment2013interaction}. Numerical spectrum in black and theoretical spectrum in red. (c) Vortex line and density iso-surface for the $k^{-3}$ case, (d) vortex coordinates $x_v(z)$ and $y_v(z)$ for the same case.}
	\label{fig-3D-KW}
\end{figure}

To compute the spectrum, we start by remeshing the line on the interval $[0,L_z]$ where $L_z$ is the simulation box size. The line is expressed as $L(z) = (l_x(z),l_y(z),z)$ and the spectrum is:
\begin{equation}
{n_{kw} = |\hat{R}(k)|^2},
\end{equation}
where $r(z) = l_x(z) + i l_y(z)$ and $\hat{R}$ is the Fourier transform of $r$. Figure \ref{fig-3D-KW} shows that the numerical spectrum is close to the imposed power law ${n_{th}}(k)$ in both cases. The script used to create the initial wave function for this case is included in the toolbox.

\clearpage
\subsection{3D quantum turbulence}\label{sec-Ex-QT}

\begin{figure}[!h]
	\centering
	\begin{subfigure}[b]{0.48\textwidth}
		\centering
		\includegraphics[width=\textwidth]{\figpath/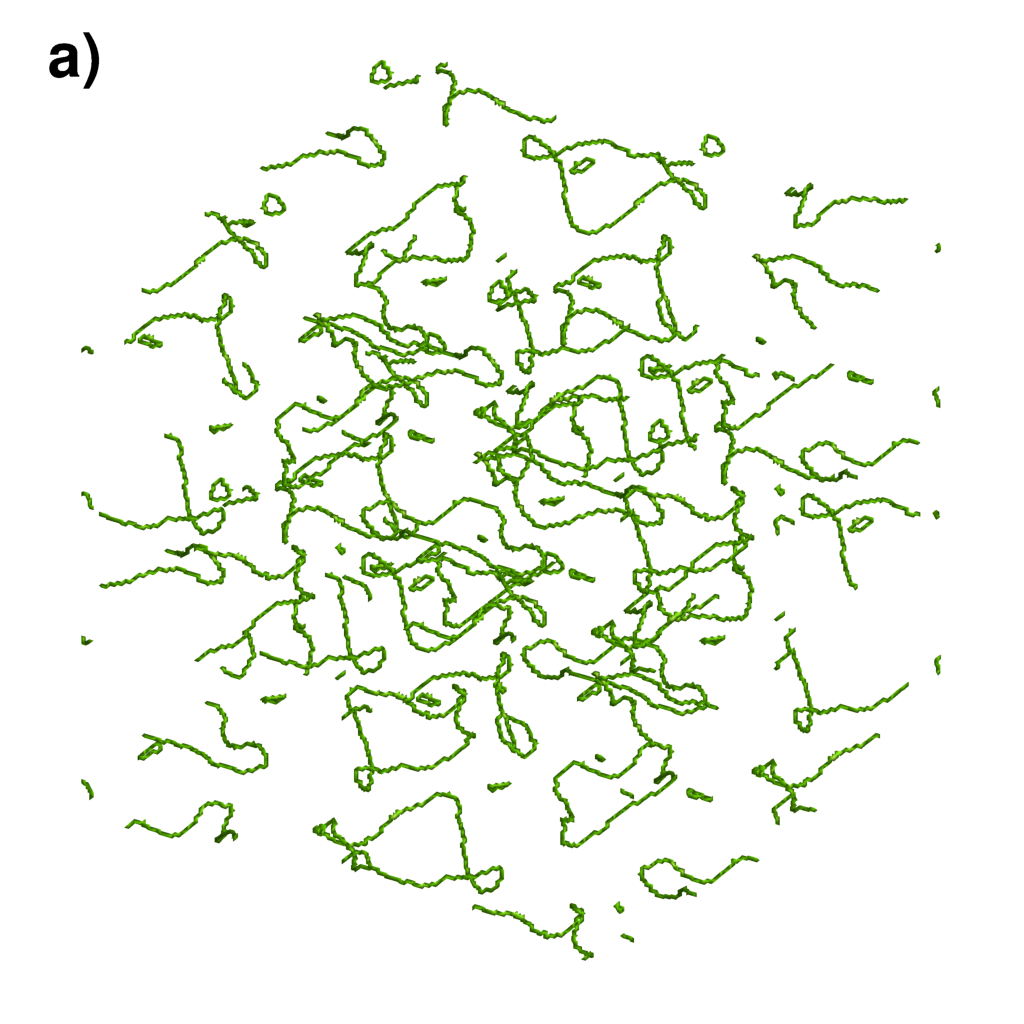}
		\label{fig:qt_a}
	\end{subfigure}
	\hfill
	\begin{subfigure}[b]{0.48\textwidth}
		\centering
		\includegraphics[width=\textwidth]{\figpath/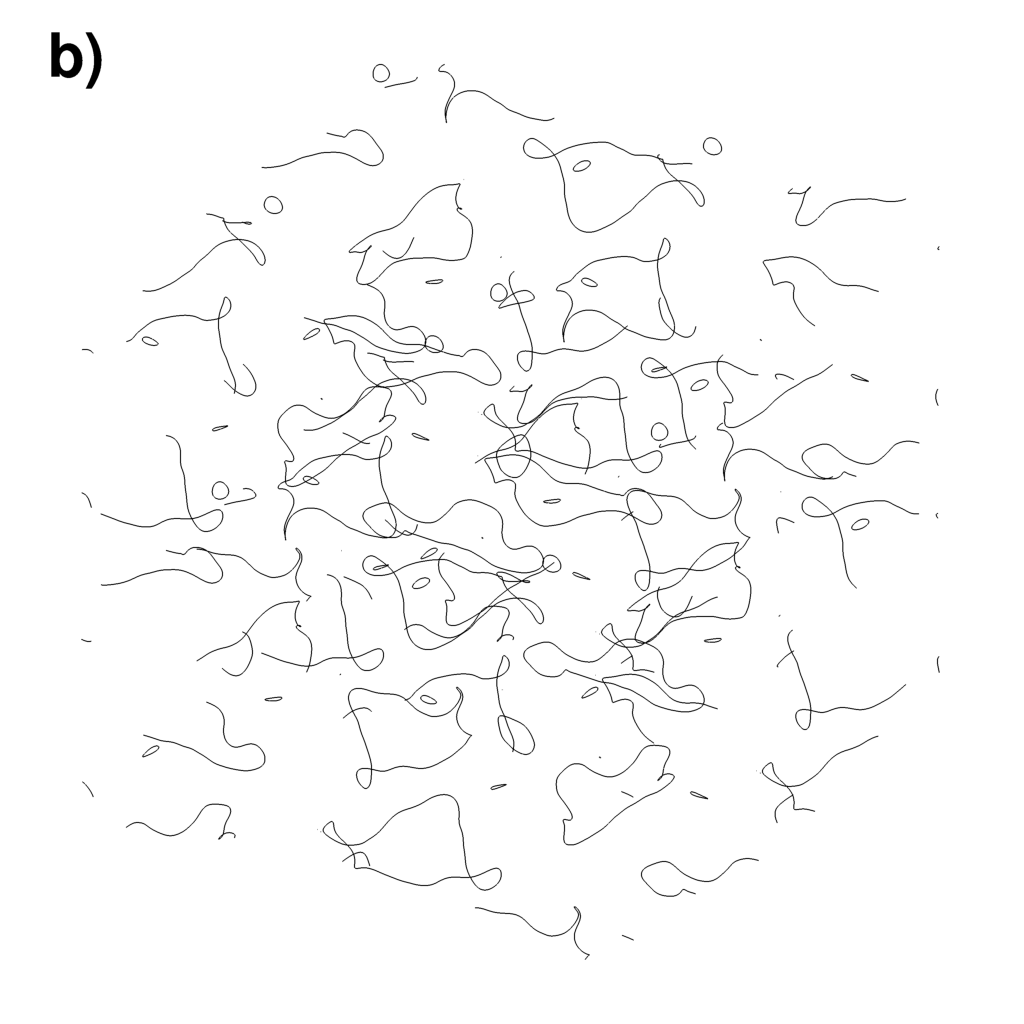}
		\label{fig:qt_b}
	\end{subfigure}
	\begin{subfigure}[b]{0.48\textwidth}
		\centering
		\includegraphics[width=\textwidth]{\figpath/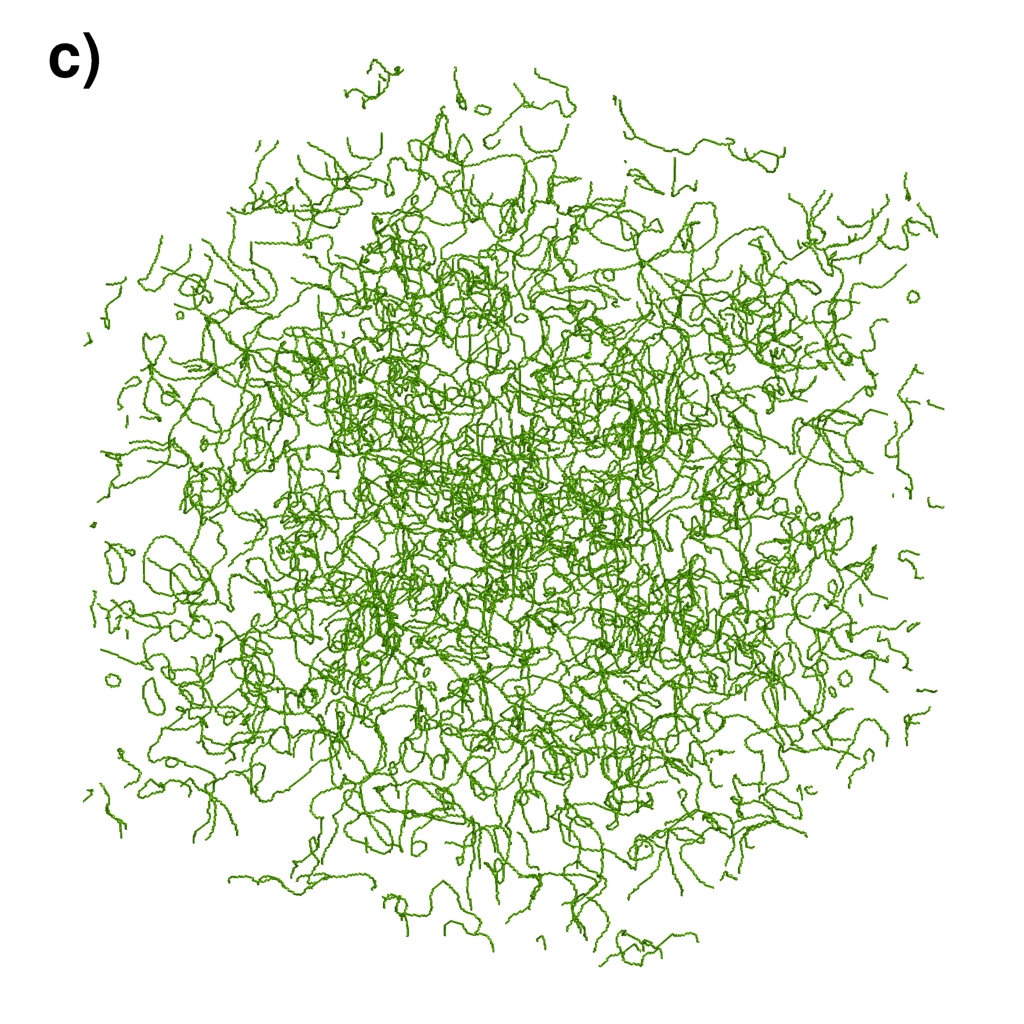}
		\label{fig:qt_c}
	\end{subfigure}
	\hfill
	\begin{subfigure}[b]{0.48\textwidth}
		\centering
		\includegraphics[width=\textwidth]{\figpath/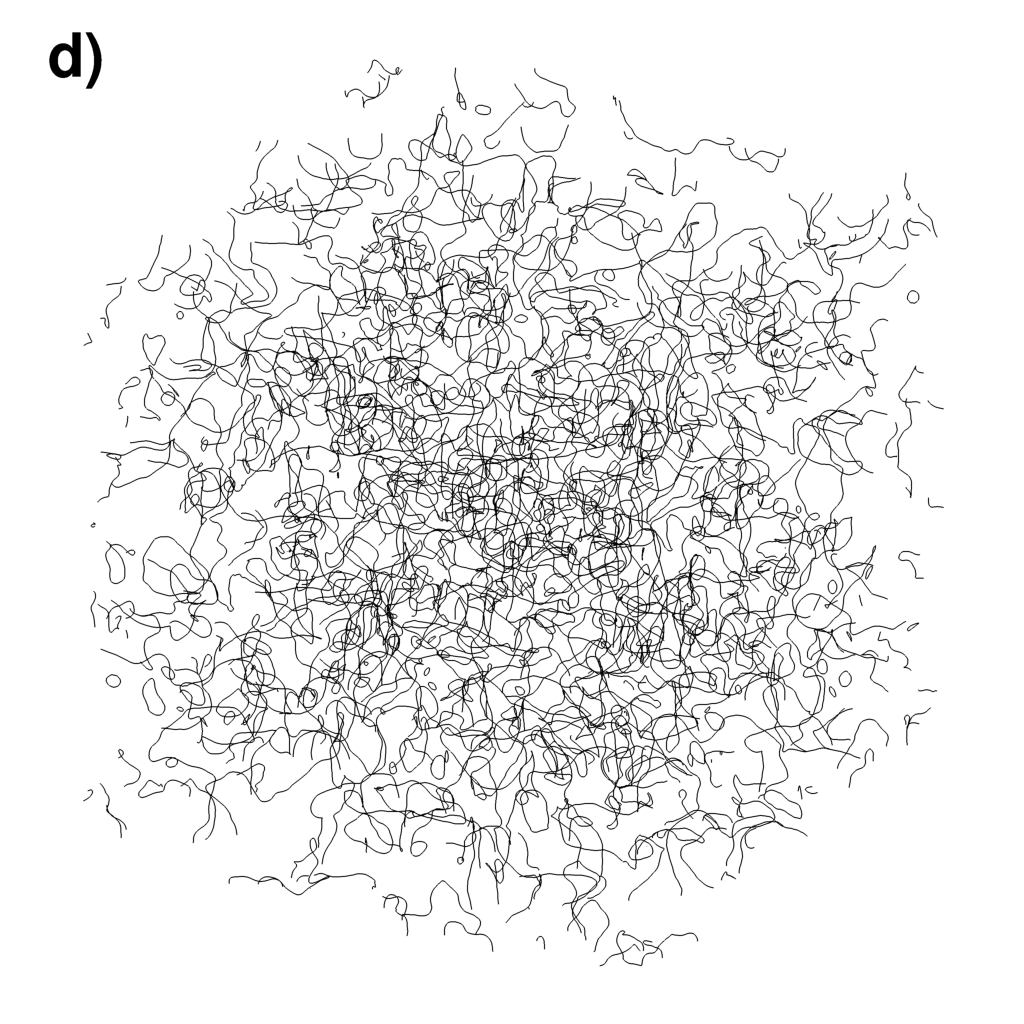}
		\label{fig:qt_d}
	\end{subfigure}
	\caption{Illustration of the vortex identification in 3D for quantum turbulence in superfluid helium. Spectral numerical simulation of the GP equation  \citep{dan-2021-CPC-QUTE}. Tetrahedrons with non-zero circulation on the left and identified vortex lines on the right for two resolutions: $128^3$ (a,b) and $256^3$ (c,d).}
	\label{fig-3D-QT}
\end{figure}

The final test case for the 3D algorithm considered  a quantum turbulence field with a large number of vortices. The wave function field was obtained from a GP spectral simulation with an initial condition containing Taylor-Green vortices \citep{dan-2021-CPC-QUTE}. Results are presented in Fig. \ref{fig-3D-QT}: $155$ vortex lines were identified from a $128^3$-grid simulation and $640$ lines were found in a $256^3$-grid simulation. The smaller vortex loops have a radius of the order of the mesh resolution.  Vortex identification is fast, as shown by computational times presented in Tab. \ref{tab-time}, despite the large number of tetrahedrons involved in the process. \cite{villois2016vortex} reported, for a similar test case, a CPU time of $6$ hours on a $64$ core cluster using MPI. As our method is local, higher resolutions can be treated by separating the simulation in small subdomains that can be analysed  separately. The vortex lines can then be linked together before the smoothing.\enlargethispage{5\baselineskip}

\subsection{Experimental dense Abrikosov lattice image}\label{sec-Ex-Coddington}

\begin{figure}[!h]
	\centering
	\begin{subfigure}[b]{0.45\textwidth}
		\centering
		\includegraphics[width=\textwidth]{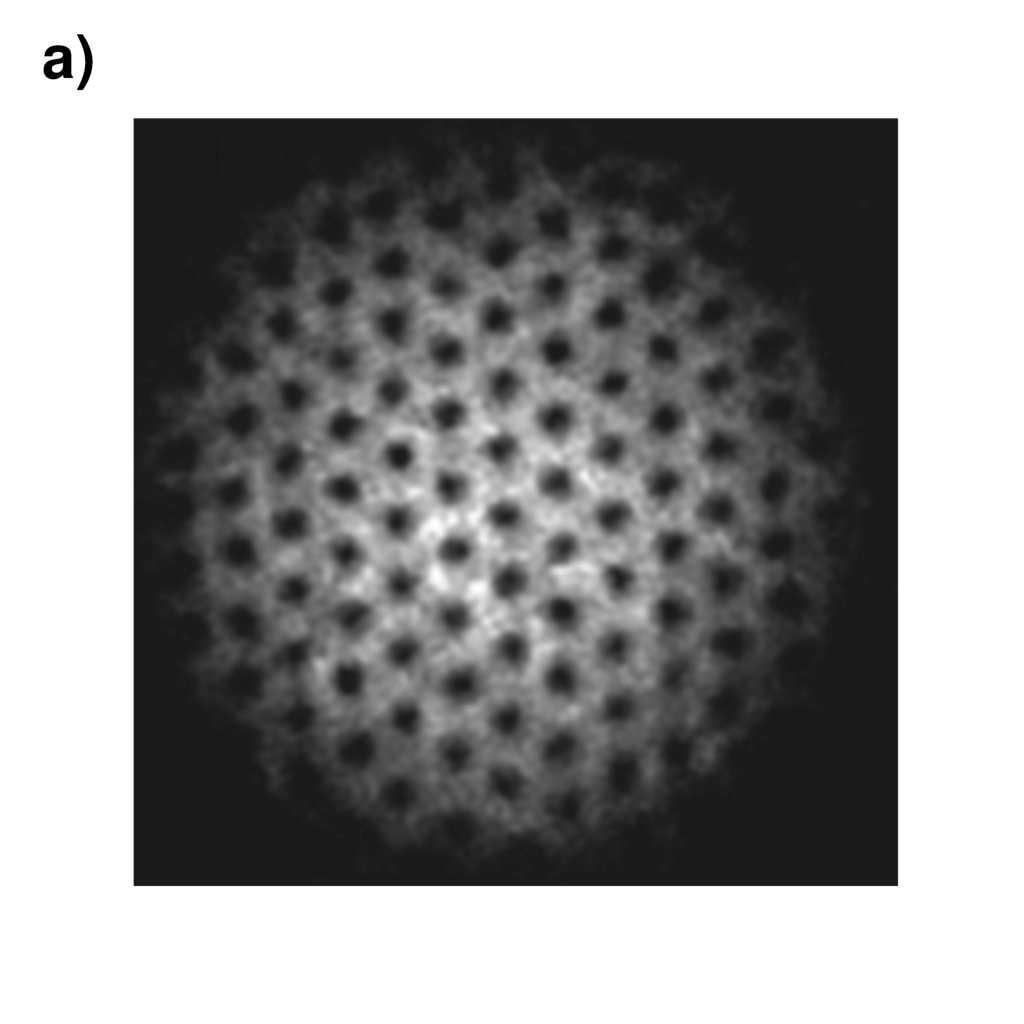}
		\label{fig:big_latt_a}
	\end{subfigure}
	\hfill
	\begin{subfigure}[b]{0.45\textwidth}
		\centering
		\includegraphics[width=\textwidth]{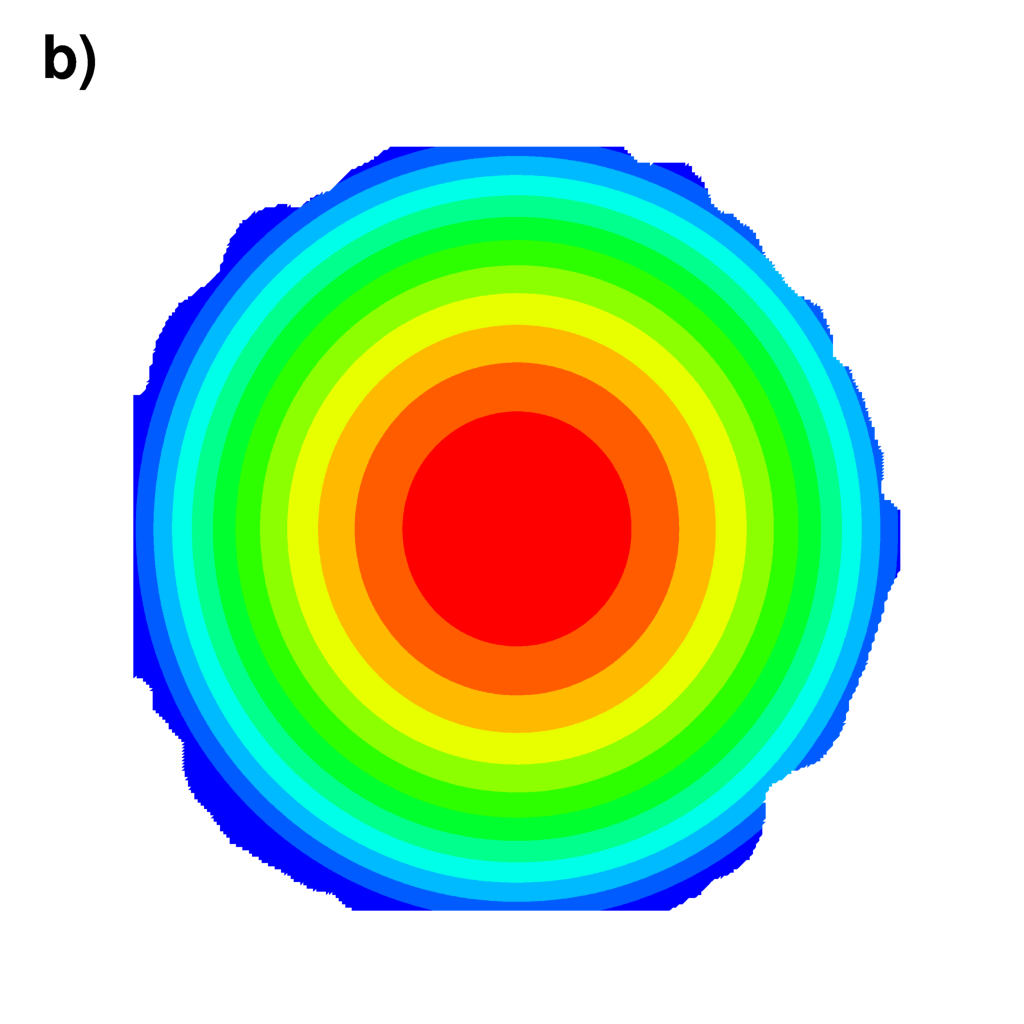}
		\label{fig:big_latt_b}
	\end{subfigure}
	\begin{subfigure}[b]{0.45\textwidth}
		\centering
		\includegraphics[width=\textwidth]{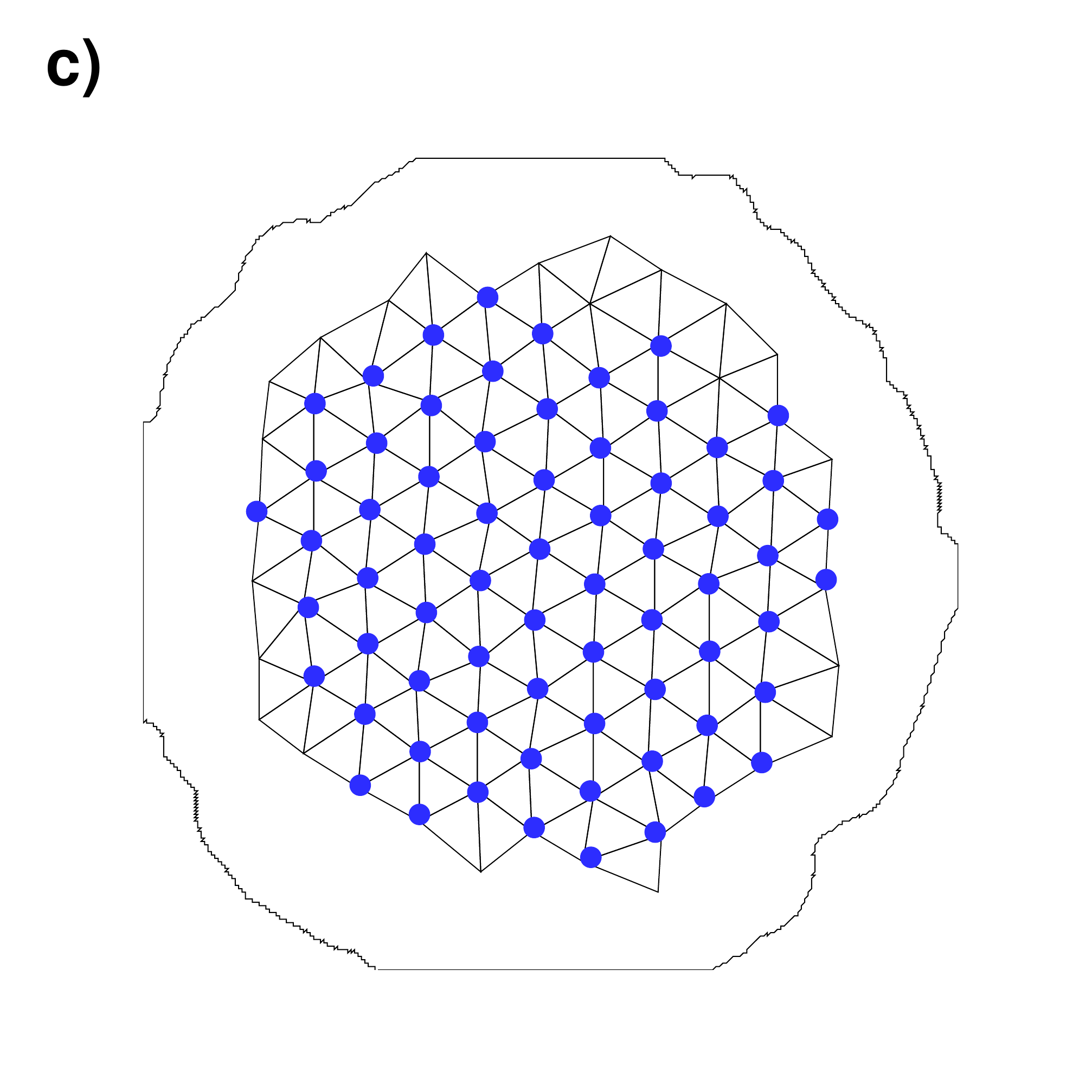}
		\label{fig:big_latt_c}
	\end{subfigure}
	\hfill
	\begin{subfigure}[b]{0.45\textwidth}
		\centering
		\includegraphics[width=\textwidth]{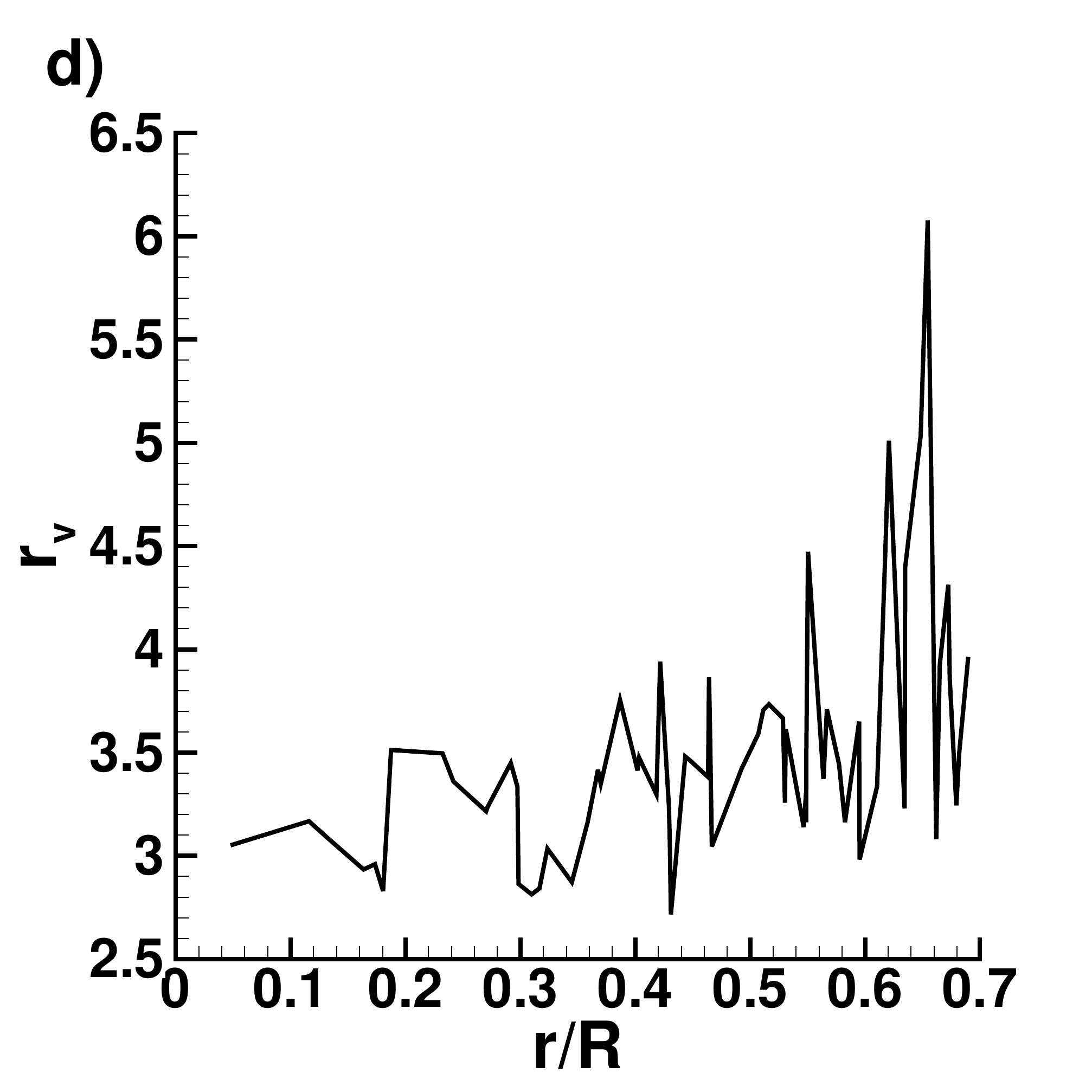}
		\label{fig:big_latt_d}
	\end{subfigure}
	\caption{Illustration of the vortex identification in an experimental image with dense Abrikosov vortex lattice \citep{BEC-physV-2001-codd}. a) Initial image, b) background density $\rho_{b}$, c) mesh of identified vortex points and Thomas-Fermi border, d) vortex radius $r_v$ as a function of the distance to the center of the BEC. In panel (c),  only vortices with high enough density contrast (and thus fitted with a Gaussian) are represented by blue circles. }
	\label{fig-2D-lattice}
\end{figure}

To complete the test presented in Sect. \ref{sec-2D-images}, we considered the more challenging test of an experimental image with a dense Abrikosov vortex lattice reported by \cite{BEC-physV-2001-codd}.
The result is shown in Fig. \ref{fig-2D-lattice}.  Vortices on the side of the condensate are not identified as they are in a very low contrast regions. The vortex lattice properties were also computed. We notice in Fig. \ref{fig-2D-lattice}(d) that the vortex radius $r_v$ increases for vortices away from the BEC center, which is the expected behaviour because of the low local atomic density. As the length scale information is not available with images, all quantities are expressed in pixel size.

\section{Description of the programs }\label{sec-desc-prog}

In this section, we first describe the architecture of the programs and the organisation of the provided files. Then we describe the input parameters and the structure of the output files for the different codes.

\subsection{Program architecture}

The \texttt{Postproc$\_$toolbox} directory is organized around three main subdirectories, \texttt{postproc\_data\_2D}, \texttt{postproc\_data\_3D} and \texttt{postproc\_image\_2D}, corresponding to the three algorithms described above: the case of 2D simulation data, that of 3D simulation data and the case of experimental images. Each of these subdirectory contains a file: {\em FFEM\_postproc\_\$case.edp} which is the main \ff script file and a folder \texttt{A\_macro} containing macros used in main code. For example, to run the vortex identification in 3D simulation data, the user can use the command  \texttt{FreeFem++ FFEM$\_$postproc$\_$data$\_$3D.edp}. Initial data fields or images for the examples presented in this paper are stored in the \texttt{INIT} folder.

The obtained solutions are saved in the folder \texttt{dircase}. Depending on the output format selected by the user,  data files are generated in specific folders for visualization with Tecplot, Paraview or Gnuplot. We also provide in the folder \texttt{Figures} ready-made layouts for Tecplot. The user can thus obtain the figures from this paper using newly generated data. More details about the output structure are given in Sect. \ref{sec-outputs}.


Before running the main scripts, it could be useful for the reader to get familiar with the initial data fields. This will help with the use of the toolbox in different programming frameworks.  For this purpose, we  included some auxiliary (optional) scripts that could be used to generate again data files for the examples presented in this paper (folder {\texttt{extra\_codes\_for\_examples}}) or to have an example of how to link an external solver (GP or similar)  to the toolbox (folder \texttt{data\_example}). More in detail:
\begin{enumerate}
	\item The directory {\texttt{extra\_codes\_for\_examples}} contains codes generating the wave function and computing vortex line properties for the  examples described in Sects. \ref{sec-Ex-bec} to \ref{sec-Ex-Kwave}:\\
	$\bullet$ {\em generate\_vortex\_knot.edp}: generates the vortex knot wave function (Sect. \ref{sec-Ex-VRknot}).\\
	$\bullet$ {\em compute\_vortex\_knot\_curvature.edp}: computes  numerical and theoretical curvatures for the vortex knot and compares them.\\
	$\bullet$ {\em generate\_sphere\_mesh.edp}: generates  a spherical mesh for the BEC example with 12 vortices (Sect. \ref{sec-Ex-bec}).\\
	$\bullet$ {\em generate\_Kelvin\_wave.edp}: generates the wave function of a vortex line deformed by a Kelvin wave  (Sect. \ref{sec-Ex-Kwave}).\\
	$\bullet$ {\em compute\_Kelvin\_spectrum.edp}: computes the Kelvin wave spectrum of the vortex line identified in the field file
	generated by the previous script.\\
	
	\item The directory \texttt{data\_example} contains example codes showing how to save numerical data and parameters in the format used in the toolbox:\\
	$\bullet$ {\em data\_example\_2D.f90}:  Fortran code computing the wave function of a 2D BEC with three vortices.\\
	$\bullet$ {\em data\_example\_3D.f90}:  Fortran code computing the wave function of a 3D BEC with a central vortex line.\\
	$\bullet$ {\em convert\_data\_2D.edp}:  \ff script converting the data created by the 2D Fortran code to the \ff format used in the toolbox.\\
	$\bullet$ {\em convert\_data\_3D.edp}:  \ff script converting the data created by the 3D Fortran code to the \ff format used in the toolbox.
\end{enumerate}

Going back to main scripts, the complete architecture of the \texttt{Postproc$\_$toolbox} main directory is the following:
\begin{enumerate}
	\item {\em FFEM\_postproc\_generate\_files.edp} is a \ff script used to automatically run preliminary or maintenance operations and generate data and mesh used in the presented examples. It is useful to run this script before testing different examples. A {\em clean} option was also added to help managing data files and results of compilation.\\
	
	\item The directory \texttt{Common\_Macros} contains one file:\\
		$\bullet$ {\em Macro$\_$common.idp}: macros used to create directories and save the results,\\
	\clearpage
	\item The directory \texttt{postproc\_data\_2D} contains codes dedicated to vortex identification in 2D numerical data  with the algorithms presented in Sect. \ref{sec-2D-data}:\\
	$\bullet$ {\em FFEM\_postproc\_data\_2D.edp}: the main \ff script.\\
	$\bullet$ \texttt{INIT}: directory containing  data and parameter files for the Abrikosov lattice and quartic potential examples described in Sects. \ref{sec-2D-data} and \ref{sec-Ex-Abrikosov}.\\
	$\bullet$ \texttt{Figures}: directory containing Tecplot layouts used to replot the figures shown in this article. The main code must be run with the associated example before opening the layout to replot the figure.\\
	$\bullet$ \texttt{A\_macro}: directory containing macros used for the vortex identification in 2D complex data.\\
	
	\item The directory \texttt{postproc\_data\_3D} contains codes dedicated to vortex identification in 3D numerical data  with the algorithms presented in Sect. \ref{sec-3D}:\\
	$\bullet$ {\em FFEM\_postproc\_data\_3D.edp}: the main \ff script.\\
	{$\bullet$ {\em FFEM\_files\_for\_Kelvin\_wave\_vortex\_knot\_lay.edp}:  script used to generate the files necessary to replot the figures in Sects. \ref{sec-Ex-VRknot} and \ref{sec-Ex-Kwave}.}\\
	$\bullet$ \texttt{INIT}:  directory containing the data and parameter files for the examples presented in Sects. \ref{sec-3D}, \ref{sec-Ex-bec} and \ref{sec-Ex-QT}: vortex rings in superfluid helium, Bose-Einstein condensate and quantum turbulence. The scripts in the {\texttt{extra\_codes\_for\_examples}} directory must be used to obtain the initial data for the examples described in Sects. \ref{sec-Ex-VRknot} and \ref{sec-Ex-Kwave}.\\
	$\bullet$ \texttt{Figures}:  directory containing Tecplot layouts used to replot the figures shown in this article. The main code must be run with the associated example before opening the layout to replot the figure. To obtain the curvature and spectrum shown in Figs. \ref{fig-3D-knot} and \ref{fig-3D-KW}, it is necessary to use the scripts in the {\texttt{extra\_codes\_for\_examples}} directory before opening the layout.\\
	$\bullet$ \texttt{A\_macro}: directory containing macros used for the vortex identification in 3D complex data.\\
	
	\item The directory \texttt{postproc\_image\_2D} contains codes dedicated to vortex identification in 2D experimental images with the algorithms presented in Sect. \ref{sec-2D-images}:\\
	$\bullet$ {\em FFEM\_postproc\_image\_2D.edp}: the main \ff script.\\
	$\bullet$ \texttt{INIT}:  directory containing the data files for the vortex lattice presented in Sects. \ref{sec-2D-images} and \ref{sec-Ex-Coddington}.\\
	$\bullet$ \texttt{Figures}: directory containing Tecplot layouts used to replot the figures shown in this article. The main code must be run with the associated example before opening the layout to replot the figure.\\
	$\bullet$ \texttt{A\_macro}: directory containing macros used for the vortex identification in 2D experimental images.\\
\end{enumerate}

\subsection{Macros and functions}

The different macros and functions used in the toolbox are:
\begin{enumerate}
	\item In \texttt{postproc\_data\_2D}, the directory \texttt{A\_macro} contains:\\
	$\bullet$ {\em M\_bec\_compute\_TF.edp}: macro computing the Thomas-Fermi density $\rho_\TF$ and condensate radius $R_\TF$.\\
	$\bullet$ {\em M\_bec\_fit\_gaussian.edp}: macro using Ipopt to fit the vortex density $\rho_{v}$ with a Gaussian function (see Eq. \eqref{eq-fit-Gauss}).\\
	$\bullet$ {\em M\_bec\_fit\_TF.edp}: macro using Ipopt to fit the Thomas-Fermi density (see Eq. \eqref{eq-JC}).\\
	$\bullet$ {\em M\_bec\_lattice\_stats.edp}: macro computing characteristics  of the vortex lattice (radius and inter-vortex spacing as a function of the distance to the center of the BEC).\\
	$\bullet$ {\em M\_bec\_read\_data.edp}: macro used to read the initial data (wave function field) and parameter files.\\
	$\bullet$ {\em M\_bec\_sort\_save\_results.edp}: macro used to save the results.\\
	$\bullet$ {\em M\_bec\_vortex\_lattice.edp}: macro building a mesh corresponding to the vortex lattice.\\
	$\bullet$ {\em M\_bec\_vortex\_regions.edp}: macro computing the vortex position (Eqs. \eqref{eq-zero-point} and \eqref{eq-circ-tri}) and building the mesh $Th_\trunc$.\\
	
	\item In \texttt{postproc\_data\_3D}, the directory \texttt{A\_macro} contains:\\
	$\bullet$ {\em M\_bec\_build\_line.edp}: macro building a vortex line following the method described in Sect. \ref{sec-3D}.\\
	$\bullet$ {\em M\_bec\_compute\_curvature.edp}: macro computing the curvature of a vortex line.\\
	$\bullet$ {\em M\_bec\_compute\_TF.edp}: macro computing the Thomas-Fermi density $\rho_\TF$ and condensate radius $R_\TF$.\\
	$\bullet$ {\em M\_bec\_read\_data.edp}: macro used to read the initial data (wave function field) and parameter files.\\
	$\bullet$ {\em M\_bec\_smooth\_line.edp}: macro smoothing a vortex line using a 5-point moving average (see Eq. \eqref{eq-smooth-line}).\\
	\clearpage
	\item In \texttt{postproc\_image\_2D}, the directory \texttt{A\_macro} contains:\\
	$\bullet$ {\em M\_bec\_fit\_gaussian.edp}: macro using Ipopt to fit the vortex density $\rho_{v}$ with a Gaussian function (see Eq. \eqref{eq-fit-Gauss}).\\
	$\bullet$ {\em M\_bec\_fit\_TF.edp}: macro using Ipopt to fit the Thomas-Fermi density (see Eq. \eqref{eq-JC}).\\
	$\bullet$ {\em M\_bec\_lattice\_stats.edp}: macro computing characteristics  of the vortex lattice (radius and inter-vortex spacing as a function of the distance to the center of the BEC).\\
	$\bullet$ {\em M\_bec\_reduceres\_smooth.edp}: macro used to smooth and lower the resolution of the initial image.\\
	$\bullet$ {\em M\_bec\_regularize\_isoline.edp}: macro used to smooth the region borders in $\rho_{iso}$.\\
	$\bullet$ {\em M\_bec\_sort\_save\_results.edp}: macro used to save the results.\\
	$\bullet$ {\em M\_bec\_vortex\_lattice.edp}: macro building a mesh corresponding to the vortex lattice.\\
	$\bullet$ {\em M\_bec\_vortex\_regions.edp}: macro computing the vortex position and building the mesh $Th_\trunc$.\\
	
	\item The {\em vortextools.cpp} file is included in the \texttt{FreeFem-sources/plugin/sec} directory for \ff versions above 4.12 and contains functions used by the previously described scripts and macros:\\
	$\bullet$ {\em smoothCurve}: function smoothing a vortex line using a 5-point moving average (see Eq. \eqref{eq-smooth-line}).\\
	$\bullet$ {\em uZero}: function computing vortices in a 3D complex field (see Eqs. \eqref{eq-zero-point} and \eqref{eq-circ-tri}); it returns a P0 function of value $1$ if the tetrahedron is crossed by a vortex line and $0$ otherwise.\\
	$\bullet$ {\em uZero2D}: function computing vortices in a 2D complex field (see Eqs. \eqref{eq-zero-point} and \eqref{eq-circ-tri}), returns a P0 function of value $1$ if the tetrahedron is crossed by a vortex line and $0$ otherwise, an array containing vortex positions and $d_{\min}$.\\
	$\bullet$ {\em ZeroLines}: function used to identify the vortex lines through the graph described in Sect. \ref{sec-3D}.\\
	$\bullet$ {\em curvatureL}: function computing the curvature of a curve described by a line 1D-mesh.\\
	$\bullet$ {\em abscisses}: function computing the arc-length along a curve.\\
	$\bullet$ {\em interpol}: function using linear interpolation to transform data from an irregular discretization to a uniform one.\\
\end{enumerate}

\subsection{Input parameters}

Parameters for the identification process must be adapted in the main code file. They differ depending on the type of data and the space dimension (2D/3D).\\
{\bf (1)} In the file \texttt{FFEM\_postproc\_data\_2D.edp}, the parameters are:
\begin{itemize}
	\item {\bf paramread}: defines the method to compute the background density: computes $\rho_\TF$ from the Gross-Pitaevskii parameters (\texttt{true}) or uses the fit presented in Eq. \eqref{eq-JC} (\texttt{false}). If \texttt{true}, a parameter file must be provided.
	\item {\bf displayplot}: controls the output information to plot. Possible values from $0$ (no plots), to {$2$} (plots data for all vortices during the fit).
	\item {{\bf save}}: {if \texttt{true}, data used in the Tecplot layouts will be saved}.
	\item {\bf iwait}: a Boolean indicating if the code must wait for user input when a plot is shown (\texttt{true}) or if it can continue (\texttt{false}).
	\item {\bf hole}: a Boolean indicating whether the condensate has a central hole as in the quartic-quadratic potential test case.
	\item{\bf fcase}: the name of the input file.
	\item {\bf dirInput}: the name of the directory where the input file is stored.
	\item {\bf dircase}: the name of the directory where the results will be stored.
\end{itemize}

{\bf (2)} In the file \texttt{FFEM\_postproc\_data\_3D.edp}, the parameters are:
\begin{itemize}
	\item {\bf BEC}: if \texttt{true}, the tetrahedrons outside the Thomas-Fermi zone will be removed. In this case, a parameter file must be given.
	\item {\bf displayplot}: controls the output information to plot. Possible values from $0$ (no plots), to {$2$} (plots data for all vortices during the fit).
	\item {{\bf save}}: if \texttt{true}, data will be saved in the \texttt{Tecplot} folder.
	\item {\bf iwait}: a Boolean indicating if the code must wait for user input when a plot is shown (\texttt{true}) or if it can continue (\texttt{false}).
	\item {\bf curv}: the vortex line curvature is computed if  \texttt{true}.
	\item {\bf npadd}: the number of points to add on each segment of the line before smoothing. The minimum is $1$ (the extremity of the segment).
	\item {\bf smooth}: a Boolean indicating whether to smooth the line or not.
	\item {\bf nsmooth}: the number of smoothing iterations.
	\item{\bf fcase}: the name of the input file.
	\item {\bf dirInput}: the name of the directory where the input file is stored.
	\item {\bf dircase}: the name of the directory where the results will be stored.
\end{itemize}  

{\bf (3)} In the file \texttt{FFEM\_postproc\_image.edp}, the parameters are:
\begin{itemize}
	\item {\bf displayplot}: controls the output information to plot. Possible values from $0$ (no plots), to {$2$} (plots data for all vortices during the fit).
	\item {{\bf save}}: if \texttt{true}, data will be saved in the \texttt{Tecplot} folder.
	\item {\bf iwait}: a Boolean indicating if the code must wait for user input when a plot is shown (\texttt{true}) or if it can continue (\texttt{false}).
	\item {\bf reduceres}: a Boolean, the image resolution is reduced when the value is \texttt{true}.
	\item{\bf fcase}: the name of the input file.
	\item {\bf dirInput}: the name of the directory where the input file is stored.
	\item {\bf dircase}: the name of the directory where the results will be stored.
	\item {\bf cutisoval}: value $c_{iso}$ used when separating the regions.
	\item {\bf cutisoval2}: a second value $c_{iso,2}$ used when separating the regions.
\end{itemize}

\subsection{Outputs}\label{sec-outputs}

When a computation starts, the \texttt{OUTPUT$\_$\$case} directory is created. It contains five folders. The \texttt{RUNPARAM} directory contains a copy of the code and data files, allowing an easy identification of each case and preparing an eventual rerun of the same case. The other folders contains different output format files of the computed solution, to be visualised with Tecplot, Paraview, Gnuplot or \ff. The content of those subfolders depends on the case and on the computation parameters:
\begin{itemize}
	\item For 2D data, the \texttt{Gnuplot} folder contains multiple files. Computed quantities are stored in the \texttt{vortex\_coord\_\$case.dat} file. The column are in order: the vortex number, the vortex coordinates $x_v$ and $y_v$, the radius, the distance from the center of the BEC and a boolean indicating if Ipopt has converged (value 0) or not (value 1) when fitting the vortex. The \texttt{sorted\_radius.dat} and \texttt{stats\_lattice.dat} contain respectively the core radius and the inter-vortex distance (second column) sorted by increasing distance to the center of the condensate (first column). The \texttt{Tecplot} folder contains files for the initial density (\texttt{rhoinit.dat}), the background density (\texttt{rhoTF.dat}), the vortex wave function on the whole domain (\texttt{rhov.dat}) and only on vortex regions (\texttt{rhovVortex.dat}),  as well as meshes for the Thomas-Fermi border of the condensate (\texttt{TF\_meshL.dat}) and the estimated vortex points (\texttt{vortex\_mesh.dat}).
	
	\item For 3D data: The smoothed vortex lines are stored as {\em .vtk} files in the \texttt{Paraview} folder, as \ff {\em .meshb} files in the \texttt{meshL} folder and as {\em .dat} files in the \texttt{Gnuplot folder} (as a succession of point coordinates $x$, $y$, $z$). The file names are \texttt{line\_} followed by {\bf fcase} and the line number. Files starting by \texttt{zeroline} in the \texttt{Gnuplot} and \texttt{Paraview} folders are the vortex lines before smoothing. The \texttt{Tecplot} folder contains the initial density (\texttt{rho\_init.dat}), { a mesh of the tetrahedrons crossed by vortex lines (\texttt{vortex\_mesh.dat})} and if it is computed, the Thomas-Fermi density (\texttt{rho\_TF.dat}). If the curvature is computed, it is saved in the Gnuplot folder as  \texttt{curv\_} followed by the line number.
	
	\item For experimental images,  the  \texttt{Gnuplot} and  \texttt{Tecplot} folders are almost the same as in the 2D data case. The only difference is that there are two meshes built from vortex points: the first one uses the points estimated from the vortex regions (\texttt{vortex\_mesh\_estimated.dat}) and the second one uses vortex positions obtained from the fit with a Gaussian (\texttt{vortex\_mesh\_fitted.dat}).
\end{itemize}

\pagebreak

\section{Summary and conclusions}\label{sec-conclusions}

The main advantage of the toolbox distributed with this paper is its ability to give fast and  accurate results for vortex identification in different configurations: 2D, 3D and experimental images. While more precise numerical methods for the identification of vortices could be set by combining local minimization and Fourier interpolation \citep{villois2016vortex,caliari2017postproc}, 
we showed that the the computation of the zeros of a P1 finite element function is a reliable method to obtain a good degree of precision while keeping the computational cost low. This approach is well suited for fields with a large number of vortices, as obtained in high resolution simulation of quantum turbulence, where local minimization would be too costly. Another advantage of the toolbox is to be adaptable to any computational framework (finite elements, spectral, finite differences) providing the wave function field. Examples how to interface the toolbox with other software are also provided.

The toolbox was created with \ff, a free and open-source software for the resolution of partial differential equations and the manipulation of finite elements. The main steps of the method are (i) the search of the zeros of the wave function (ii) the delimitation of the vortex region (in 2D) or the vortex line (in 3D) (iii) the computation of vortex characteristics. The numerical code was validated against a wide range of test cases available in the literature for both 2D and 3D configurations. 

To facilitate its use, the toolbox is provided with separate folders containing all the necessary files (parameters, restart files) to run all the cases described in the paper. Ready-made scripts and layouts allow the user to generate the figures presented in this paper with newly generated data after running the programs. As a consequence,the present toolbox can be easily tested and modified for the computation of other diagnostics or the identification of vortices in other types of systems (\eg vortices in superconductors described by a Ginzburg-Landau wave function field).

\section*{Acknowledgements}

The authors acknowledge financial support from the French ANR grant ANR-18-CE46-0013 QUTE-HPC. Part of this work used computational resources provided by IDRIS (Institut du d{\'e}veloppement et des ressources en informatique scientifique) and CRIANN (Centre R{\'e}gional Informatique et d'Applications Num{\'e}riques de Normandie).


\section*{Bibliography}


\end{document}